\documentclass[a4paper,11pt]{article}
\pdfoutput=1 

\usepackage{jheppub} 
\usepackage[english]{babel}

\usepackage[nottoc]{tocbibind}

\usepackage[T1]{fontenc} 
\usepackage{amsmath}
\usepackage{amsfonts}
\usepackage{amssymb}
\usepackage{hepunits}
\usepackage[font=small,skip=0pt]{caption}
\usepackage{float}
\usepackage[utf8]{inputenc}
\usepackage[colorinlistoftodos]{todonotes}
\usepackage[nottoc]{tocbibind}

\title{\boldmath Constraints on Relic Magnetic Black Holes }

\author{Melissa~ D.~ Diamond,\note{corresponding author} David~ E.~ Kaplan  }

\affiliation{Department of Physics $\&$ Astronomy,\\The Johns Hopkins University, Baltimore, MD 21218, USA\\
Department of Physics\\}

\emailAdd{mdiamon8@jhu.edu,david.kaplan@jhu.edu}

\abstract{We present current direct and astrophysical limits on the cosmological abundance of black holes with extremal magnetic charge.  Such black holes do not Hawking radiate, allowing those normally too light to survive to the present to do so.   The dominant constraints come from white dwarf destruction for low and intermediate masses ($2\times10^{-5}$ g -- $4\times 10^{12}$ g) and Galactic gas cloud heating for heavier masses ($>4\times 10^{12}$g).  Extremal magnetic black holes may catalyze proton decay. We derive robust limits -- independent of the catalysis cross section -- from the effect this has on white dwarfs.  We discuss other bounds from neutron star heating, solar neutrino production, binary formation and annihilation into gamma-rays, and  magnetic field destruction. Stable magnetically charged black holes can assist in the formation of neutron star mass black holes.}

\begin{document} 
\maketitle
\flushbottom

\section{Introduction}
A simple,  well-motivated extension of the Standard Model of particle physics (and Maxwell's equations)  is the inclusion of magnetic monopoles.  Magnetic monopoles are a natural consequence of grand-unified theories (GUTs)  \cite{1974NuPhB..79..276T,Polyakov:1974ek},  have been searched for by a number of different experiments  \cite{Ambrosio:2002qq,PhysRevLett.66.1951,Balestra_2008}, and were an original motivation for cosmic inflation  \cite{PhysRevD.23.347,Preskill:1979zi}.

One possible manifestation of magnetic charges in our universe is magnetically charged black holes.  Primordial black holes could have formed in the early Universe from, for example, large-amplitude, short-wavelength density perturbations or early phase transitions  \cite{Carr:1974nx,PhysRevLett.44.631}.  If monopoles existed (or perhaps dominated) at that time, one might expect $\sqrt{N}$ fluctuations to leave primordial black holes with magnetic charges.  Light enough black holes could eventually Hawking radiate until they become extremal, leaving a magnetically charged remnant today, a fate different from that of small uncharged black holes.  This is touched upon in Ref.\cite{Araya_2021}. Additionally near Planck mass primordial black holes could acquire magnetic charge by spontaneously emitting magnetic monopoles, leading their charge to fluctuate and perhaps get caught at a stable extremal value.  This formation process is explored for extremal electrically charged black holes in Ref.\cite{Lehmann:2019zgt}.  More exotic formation scenarios may exist, but we will not address those here.   EMBHs are interesting because they are not expected to Hawking radiate, making them stable dark matter candidates at masses ruled out for uncharged black holes.  Their large mass compared to that predicted for magnetic monopole also allows many EMBHs to avoid the constraints on magnetic monopoles, effectively providing a way to ``hide'' magnetic charge.  Understanding the behavior of EMBHs gives insight into other exotic heavy magnetic objects, such as some higher dimensional black holes \cite{Bah:2020ogh}.

In this paper, we put constraints on the abundance of primordial black holes with magnetic charge.   We focus on extremal magnetic black holes (EMBHs), but point out the applicability and strength of the bounds for magnetic black holes (MBHs) heavier than $10^{15}$g with non-extremal charges.  The dominant bounds come mainly from two sources -- Galactic gas cloud heating at large masses ($>4\times 10^{12}$g) and destruction of white dwarfs (WDs) at lower masses.  We present novel constraints from gamma-ray emission for EMBHs at low masses, though they  are never  dominant.  In addition, we touch upon bounds from the destruction of magnetic fields and direct searches at MACRO.

Magnetic monopoles derived from some GUTs are predicted to catalyze proton decay with significant cross sections \cite{Callan:1983nx,Rubakov:1982fp}. The bounds on EMBHs change significantly when one assumes catalyzed proton decay in stellar objects -- neutron stars, white dwarfs, and the  Sun -- due to heating and neutrino emission.  We scan over cross sections for this catalysis to generate robust bounds on relic abundances of EMBHs, independent of the size of this effect.  The dynamics of nucleon decay near EMBHs are highly non-trivial, and we leave a detailed analysis for future work.

This paper is organized as follows.  In Section \ref{sec:Theory}, we briefly discuss the relevant physical properties of EMBHs.  In Section \ref{sec:BoundsNoCat}, we describe bounds due to the annihilation of binary EMBHs, the heating of the intergalactic medium, the destruction of WDs, destruction of large scale magnetic fields, and non-detection of magnetic charges neglecting effects from catalysis of proton decay.  In Section \ref{sec:BoundsCat}, we include the catalysis of proton decay for all possible cross sections.  This weakens the bounds in some mass regimes by, for example, preventing the destruction of WDs.  For illustration, we also show, for a fixed catalysis cross section,  bounds  which arise from neutron star (NS) heating and proton decay in the Sun.  In Section \ref{sec:nonextremal},  we explore bounds on MBHs heavy enough to be stable over cosmological timescales even when sub-extremal.  Section \ref{sec:Conclusions} includes comments on how to generalize the bounds on EMBH abundance when EMBHs have a non-monochromatic mass spectrum, how EMBHs may seed large scale magnetic fields.  Section \ref{sec:Conclusion} concludes.  We compare our constraints with those from Refs.\cite{bai2020phenomenology,ghosh2020astrophysical} in Figs.\ref{fig:non-cat} and \ref{fig:catminlim}.  Throughout this paper, unless units are explicitly specified otherwise, we take $c=\hbar=k =1$ and $\epsilon_0 = \frac{1}{4\pi}$.

\section{Extremal Magnetic Black Holes (EMBHs)}
\label{sec:Theory}

In general relativity, MBHs are described by the Reissner-Nordstrom metric, one that is spherically symmetric and carries an outer event horizon and an inner Cauchy horizon.  The extreme limit of the magnetic charge is one in which these horizons merge, producing the metric:
\begin{equation}
    ds^2 = - \left(1 - \frac{GM}{r}\right)^2 dt^2 + \left(1 - \frac{GM}{r}\right)^{-2} dr^2 + r^2 (d\theta^2 + \sin^2\theta d\phi^2),
\end{equation}
where $M$ is the ADM (asymptotic) mass of the black hole and $G$ is Newton's constant. The EMBH has, in natural units, a magnetic charge  
\begin{equation}
\label{chargemass}
    Q  = \frac{M}{m_{P}},
\end{equation}
where $m_{P} = G^{-1/2} = 1.22\times10^{19}$ GeV is the Planck mass.  In these units, $Q = 10^5$ corresponds to $M\sim 1$ g.   The minimum value for $Q$ is set by the Dirac quantization $Q_{min}=\frac{1}{2e}\sim6$, where $e$ is the electron charge  \cite{1931RSPSA.133...60D}. Note that this is a different definition of Q than is used in other works \cite{Maldacena:2020skw, bai2020phenomenology} on EMBHs. 

EMBHs have some interesting and unresolved properties.  Being extremal, they will not Hawking radiate  \cite{1974Natur.248...30H}.  Uncharged black holes with $M\lesssim 5\times10^{14}$g are predicted to decay within the lifetime of the Universe. EMBHs are expected to be stable and could, in principle, be cosmological relics.\footnote{Recent work  \cite{Maldacena:2020skw} suggests that EMBHs may have finite lifetimes if they decay into magnetic monopoles or smaller EMBHs.  The lifetime is model-dependent and sensitive to properties of the magnetic monopole, so we will not consider it here.}  Ref.\cite{Maldacena:2020skw} indicates that there are large regions around EMBHs where the magnetic fields are strong enough to condense electroweak bosons and restore the Higgs field minimum to its origin, making Standard Model fermions massless.  Interestingly, one would also expect the sphaleron barrier for baryon number violation to disappear \cite{ho2020electroweak} in this region.  In addition, the interior of non-extremal magnetically charged black holes will have an inner Cauchy horizon with known instabilities  \cite{PhysRevLett.63.1663}.  In general relativity, the interior metric can only be extended in physically nonsensical ways (requiring the existence of an infinite number of other universes).  It is thus expected that new physics lives at this horizon, and in the extremal limit, this would be at $r=GM$.  The physics at this surface, for example, could involve the Planckian density \cite{Kaplan_2019} and violate baryon number in a significant way or source other fields (black hole `hair').  Thus significant model dependence dictates the physics at close ranges.

A possibly important feature of EMBHs is their potential for catalyzing proton decay, akin to that in GUT-induced magnetic monopoles  \cite{Callan:1983nx,Rubakov:1982fp}.  It is a non-trivial question here -- not only is there model dependence with respect to the properties of the core of the EMBH, but also the path to the core for a charged particle, which will involve gauge-boson interactions at medium and long distances and gravitational and (potentially) black-hole hair interactions at short distances\footnote{For example, an EMBH should gravitationally repel  particles with a large charge to mass ratio ({\it e.g.}, the proton), as the end result would be super-extremal.  This would also be true of particles with spin.}. 
A full analysis for specific models of EMBHs would be interesting.  However, to put a robust bound on these objects, we will allow the proton catalysis cross section to be a free parameter (within reason), and will see it is possible to put more  {\it model-independent} limits from various astrophysical phenomena.

A number of other interesting properties of EMBHs were explored in  Ref.\cite{Maldacena:2020skw}.  One is that charged Fermions near an EMBH will experience radically different dynamics.  In regions where the magnetic field is larger than the fermion's squared mass, the low-lying energy quantum states are Landau levels and should correspond to two-dimensional fermion states which move radially and have degeneracy of order $Q$.  This is predicted to enhance the Hawking decay rate of non-extremal black holes by a factor of $Q$.  This will have some effect on the discussion of bounds on merger rates.

\section{Bounds on EMBHs without Catalysis}
\label{sec:BoundsNoCat}

\begin{figure}
    \centering
    \includegraphics[scale=0.45]{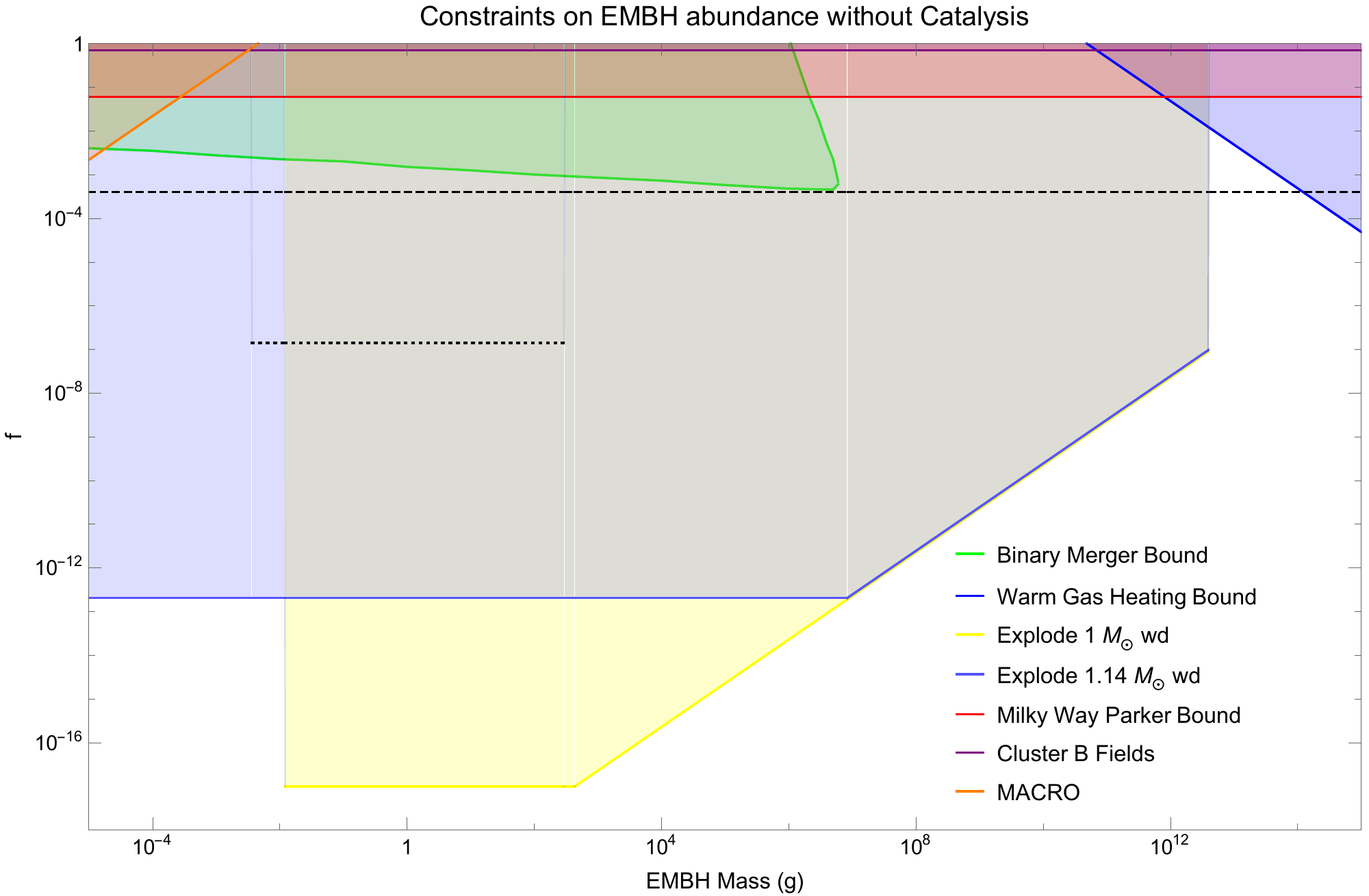}
    \caption{These are the bounds on the EMBH abundance, $f$, if nucleon catalysis effects, described in Section \ref{sec:BoundsCat}, are neglected.  The green region is ruled out by gamma-rays emitted from binary mergers, (see Section \ref{sec:binarymerger}). The yellow and light blue regions are respectively ruled out by the existence of 1 Gyr old non-magnetic 1 M$_{\odot}$ WDs and by WD J0551+4135 which EMBHs can destroy (see Section \ref{sec:wdann}). The dark blue region is ruled out by EMBHs overheating warm ionized gas clouds in the Milky Way (see \ref{warmgas}).  The red and purple regions are ruled out by EMBH interaction with large scale magnetic fields (see Section \ref{sec:destruction of Mfield}).  The orange region is ruled out by monopole searches by MACRO, which is described in Section \ref{sec:MACROnocat}.  For comparison, we show EMBH bounds set by Bai et al. in Ref.\cite{bai2020phenomenology} in black.  The dashed bound comes from the Parker effect in the Andromeda Galaxy (see Section \ref{sec:destruction of Mfield}), and the dotted one comes from EMBH annihilation in the Sun producing excess neutrinos (see Section \ref{sec:inSun}).  We cut off the latter bound at $M>300$ g because convection in the Sun can suppress the annihilation rate for these EMBHs.}
    \label{fig:non-cat}
\end{figure}

Relic EMBHs are heavy magnetic charges  expected to be non-relativistic at all relevant times and follow the dark matter distribution once they decouple from the background plasma.  Their physical effects include gas heating, WD destruction, magnetic field destruction, and high-energy particle emission from annihilation.  We describe the bounds from these effects in the subsections below and summarize them as constraints on the average relic density of EMBHs as a fraction of the average  dark matter energy density, $f\equiv{\rho_{bh}/\rho_{dm}}$.  We take the dark matter density in the Milky Way to be $\rho_{dm}^{MW} = 0.3 $ GeV cm$^{-3}$.  In this section, we are ignoring the possibility of proton-decay catalysis.
We plot the bounds (without catalysis)  for a range of masses in Fig.\ref{fig:non-cat}.

\subsection{Gas Cloud Heating}
\label{warmgas}
When an EMBH passes through an ionized plasma, it loses energy by exciting long range Eddy currents.  This heats the plasma via the kinetic energy of the accelerated electrons. We constrain $f$ by requiring that EMBHs not change the observed temperature of clouds of warm ionized medium (WIM) in the Milky Way.  The heating rate from friction must be less than the average cooling rate determined in a survey of the WIM  \cite{Lehner:2004ty}.

The friction  between a magnetic monopole and a plasma is described in  Ref.\cite{1985ApJ...290...21M}. We adapt it here for an EMBH with charge $Q$. 
\begin{equation}
\label{eqn:friction}
    \left(\frac{dE}{dx}\right)_{\!\! p} = -\frac{16 \pi^{1/2}e^2n_e}{3\sqrt{2T m_e}}Q^2 v \left[\text{ln}\left( 4\pi n_e \lambda_D^2l\right)+\frac{2}{3}\right],
\end{equation}
where $n_e$, $m_e$ are the electron number density and mass, respectively, $T$ is the gas temperature, $v$ is the velocity of the EMBH, $\lambda_D=\sqrt{\frac{T}{4\pi n_e e^2}}$ is the plasma Debye length, and $l = \left(\frac{2T}{\pi m_e}\right)^{1/4}\frac{1}{v^{1/2}\omega_p}$ is the attenuation length of the plasma with plasma frequency $\omega_p = \sqrt{\frac{4\pi e^2 n_e}{m_e}}$.  Ionized gas clouds observed on this cooling survey \cite{Lehner:2004ty} have an average $n_e \sim 0.08 \text{ cm}^{-3}$ and  $T\sim 6000\text{ K}$.  Using Eq.\eqref{eqn:friction} we find that a single EMBH moving through at virial velocity $v\sim 10^{-3}$ will deposit energy into the plasma at a rate of
\begin{equation}
\label{eqn:Elossgas}
    \frac{dE}{dt} = 9\times10^{6} \left(\frac{M}{10^9\text{ g}}\right)^2 \text{erg s}^{-1}.
\end{equation}
Unregulated by cooling, this will significantly change the temperature and behavior of the gas within $\delta t \sim \frac{T}{dT/dt}\sim 1\text{ Myr} \left(\frac{10^9\text{g}}{M}\right)
\frac{1}{f}$.  A wide range of EMBHs can, therefore, change the properties of the WIM in the Milky Way in a time comparable to that of other less exotic heating processes \cite{1977ApJ...218..148M}.

The temperature of WIM clouds is regulated by a variety of heating and cooling processes.  The primary one is electron and hydrogen atom collisions with singly ionized carbon atoms.  These excite fine structure transitions in the ground state of carbon atoms from the $^2P_{1/2}$ to $^2P_{3/2}$ state.  Observations of infrared emission from de-excitations of these carbon atoms are used to estimate the cooling rate in ionized gas clouds  \cite{Lehner:2004ty}, which is sensitive to the temperature.  The heating rate of the gas is estimated from  and should not exceed the observed cooling rate \cite{Lehner:2004ty}.  Doing so would disturb the observed dynamics of the gas, changing both its temperature and cooling rate. 

Cooling rates for a variety of WIM clouds are reported in  Ref.\cite{Lehner:2004ty} and grouped by the velocity of the cloud.  We compare the EMBH gas heating rate to the cooling rate for low velocity clouds because this gives the strongest bounds.\footnote{In principle, a somewhat stronger bound can be found using high velocity gas clouds.  However there is significant uncertainty around $n_e$ in these gas clouds, making it difficult to compare the cooling rate to the heating rate from EMBH. } The average cooling rate  is $(dE/dt)\sim n_H\times  10^{-25.65\pm^{0.11}_{0.15}}\text{ erg s}^{-1}$, where $n_H
\sim0.25$ cm$^{-3}$ is the number density of hydrogen atoms in a typical cloud  \cite{Lehner:2004ty,1977ApJ...218..148M}.  The rate is given with  $3\sigma$ uncertainties. Requiring that EMBH WIM heating not exceed the $3\sigma$ upper limit in the cooling rate gives the following constraint:

\begin{equation}
\label{gaslim}
    f \lesssim 1.6 \times \left(\frac{M}{10^9\text{ g}}\right)^{-1} \left(\frac{n_H}{0.25\text{ cm}^{-3}}\right).
\end{equation}
This bound is only sensible if there are enough EMBHs, their density in the Galaxy being $\sim 10^{31} f \left(M/\text{g}\right)^{-1} $ pc$^{-3}$, in a warm gas cloud to heat it. Warm ionized gas clouds have a typical radius of $\sim$2 pc and would thus be well populated by EMBHs with masses at least up to $10^{15}$g \cite{1977ApJ...218..148M}.  Gas heating effects from EMBHs with $M>10^{15}$ g is discussed in Section \ref{sec: Gas Heating}.  The above constraint is plotted in Fig.\ref{fig:non-cat}.

\subsection{White Dwarf Destruction}
\label{sec:wdann}

``Small'' energy injections in the WD core of a Carbon-Oxygen (CO) WD can initiate runaway fusion, which leads to Type Ia supernovae \cite{1992ApJ...396..649T}.   For example, depositing $\sim6\times10^{21}$ GeV within a region of radius of $\sim 6\times 10^{-3}$ cm in less than $\sim 3 \times 10^{-12}$ s leads to runaway fusion and supernovae in WDs with masses above $1 M_\odot$  \cite{Janish:2019nkk,Fedderke_2020,1992ApJ...396..649T}.  EMBHs collect inside   WDs, and can eventually merge with each other and annihilate.  The merger event releases a large amount of electromagnetic (EM) energy, typically more than enough to destroy a WD.   We use this phenomenon and the observation of old (>1 Gyr) WDs in the Galaxy to constrain the EMBH abundance.  

The energy needed to destroy a WD depends on the its mass, with lighter WDs requiring larger energy deposits.  Not all EMBHs are heavy enough to supply the energy needed to destroy all WDs.  We account for this by considering bounds from two different sets of WDs: a collection of known non-magnetic $1 M_{\odot}$ WDs, and WD J0551+4135, a $1.14 M_{\odot}$  WD thought to have formed from the collision of two smaller CO WDs \cite{Hollands_2020}. 
We use $1 M_{\odot}$  WDs for multiple reasons. Hundreds such WDs have been observed  \cite{2017ASPC..509....3D}.  Their average age is 1.9 Gyr, though some ages over $\sim$ 10 Gyr  \cite{2017ASPC..509....3D}. Even allowing for some uncertainty in their cooling ages, more than half would not exist if EMBHs could destroy them within 1 Gyr.  WDs with starting masses above $1.05 M_{\odot}$ are expected to be primarily composed of Oxygen and Neon, instead of Carbon and Oxygen \cite{2007A&A...476..893S}, and require much larger energy depositions to trigger runaway fusion \cite{1992ApJ...396..649T}. These do not give useful bounds.   $1 M_{\odot}$  is about as heavy (having the smallest trigger energy) as CO WDs can initially form.

We also consider bounds from WD J0551+4135, a rare CO WD heavier than $1.05 M_{\odot}$ thought to have formed from the collision of two smaller CO WDs just over 1 Gyr ago \cite{Hollands_2020}.  Its high mass of $1.14 M_{\odot}$ makes its trigger energy low enough that EMBHs too light to trigger runaway fusion in  $1 M_{\odot}$ WDs can destroy it.  Key physical properties of these WDs used for calculations are shown in Table \ref{tab:table1}.
\begin{table}[htp]
\begin{center}
\begin{tabular}{|c|cc|}\hline $M_{wd}$ ($M_\odot$) & 1 & 1.14 \\ \hline $R_{wd}$ (km) & 5600 & 4600 \\${\rho_{c}}$ (g/cm$^3$) & $2.9\times 10^7$ & $7\times 10^7$ \\$E_T$ (GeV) & $7\times 10^{21}$ & $3\times 10^{19}$ \\$\tau_{diff}$ (s) & $3\times 10^{-12}$ & $6\times 10^{-13}$ \\$B_c$ (kG) & 1 & 50 \\$M_{min}$ (g) & 0.012 & $2\times 10^{-5}$ \\$M_{Th}$ (g) & 2600 & $5\times 10^{7}$ \\$M_{max}$ (g) & $4\times 10^{12}$ & $4\times 10^{12}$ \\ \hline \end{tabular} 
\caption{\\ Above are the relevant properties of the WDs used to derive bounds in this work.  The radius is $R_{wd}$, $\rho_c$ is the density at the core, $E_T$ is the threshold energy needed to initiate runaway fusion, $\tau_{diff}$ is the diffusion time, and $B_c$ is the estimated magnetic field at the core.  The EMBH masses $M_{min}$, $M_{Th}$, and $M_{max}$ are the minimum mass needed to  trigger runaway fusion, the threshold mass required to overcome the WD core magnetic field, and the maximum mass that can sink to the core in less than a Gyr, respectively. The first column is for generic non-magnetic 1 $M_{\odot}$ CO WDs and the second is for WD J0551+4135.    The radii, densities, and diffusion times are derived using WD profiles from Ref.\cite{Fedderke_2020}.}
\label{tab:table1}
\end{center}

\end{table}
Both sets of WDs accumulate EMBHs, which sink to the core, merge and initiate runaway fusion.

WDs are composed of a degenerate Fermi gas of electrons and an ion plasma of nuclei.  Friction with the Fermi gas and ion plasma traps traversing EMBHs.   The stopping power of a degenerate Fermi gas at zero temperature for a monopole is given in  Ref.\cite{PhysRevD.26.2347}.  Adapting this for an EMBH, we find:

\begin{equation}
\label{WDfric}
    \frac{dE}{dx} = \frac{GM^2\omega_p^2v}{v_{f}}\left[\text{ln}\frac{4 E_f}{\nu_{\sigma}^{ei}}-\frac{1}{2}\right],
\end{equation}
where $v_f$ is the Fermi velocity, $E_f$ is the Fermi energy, 
and $\nu_{\sigma}^{ei}\sim 10^{17}$ s$^{-1}$ is the electron ion scattering frequency \cite{Potekhin:1999yv} at relevant WD densities and temperatures \cite{DAntona:1990kji} (we use  $10^6$ g/cm$^3$ and $10^{6.8}$ K respectively).   For these parameters the energy loss rate is
\begin{equation}
\label{eqn:easyfric}
    \frac{dE}{dx} \simeq 1.1\times 10^{18}\times  \left(\frac{\rho_{wd}}{10^6\text{g cm}^{-3}}\right)\left(\frac{M}{g}\right)^2 \left(\frac{v}{2.3\times 10^{-2}}\right) {\rm\, GeV/km} ,
\end{equation}
where $\rho_{wd}$ is the average white dwarf density.  The velocity is scaled to the escape velocity at the surface of a $1 M_{\odot}$ WD, $v_{esc}\sim2.3\times10^{-2}$, for illustrative purposes. 

The above description of the friction is appropriate when the energy gained by electrons passing the EMBH, $\sim pv$,  is large compared to the temperature of the Fermi gas. Here, $p$ is the momentum of passing electrons.  When the energy exchanged is small compared to the temperature of the Fermi gas, finite temperature effects must be considered.  The stopping power for a magnetic charge in a Fermi gas with a finite temperature is shown in Appendix \ref{sec:appfermi} to be
\begin{equation}
\label{fermifric}
    \left(\frac{dE}{dx}\right)_{\!\! f}=\frac{Ge^2\sqrt{2m_e}M^2}{\pi^2}\text{Log}\left[\frac{2E_f}{\nu_{\sigma}^{ei}}\right]\int_{-1}^{1}\int_{-1}^{1}\int_0^{\infty}\sqrt{E'}\times f(E')\left(1-f(E'+\Delta E')\right)
    dx dy dE'
\end{equation} where
\begin{equation}
    f(E') = \frac{1}{\text{exp}\left(\frac{E'-E_f}{T}\right)+1},
\end{equation}
 is the fraction of states occupied for a given electron energy $E'$, and
\begin{equation}
    \Delta E' = pv(x-y)
\end{equation}
is the energy gained by a passing electron.  $E_f$ is the Fermi energy of the gas, and $x$ and $y$ are respectively the angles between the electron's incoming and outgoing momentum and $v$.  The total friction in a WD can be estimated by combining the contributions from \eqref{fermifric} and \eqref{eqn:friction} for a plasma of carbon nuclei.  

The lightest EMBHs we consider, $M\sim 2\times 10^{-4} {\rm \,g} \sim 10^{20} {\rm\, GeV}$, have kinetic energies within the Galaxy of $\sim 6\times 10^{13} {\rm\, GeV}$.  Using Eq.\eqref{eqn:easyfric} to make a simplified estimate of the energy losses, we find that the lightest EMBHs get trapped after traversing about $\sim 1000$ km inside a WD.  As heavier EMBHs lose kinetic energy over even shorter distances, it is clear that every one which passes through a WD becomes gravitationally bound to it.  WDs accumulate EMBHs at a Sommerfeld enhanced rate:
\begin{equation}
\label{interaction rate}
    \Gamma_{wd} \simeq \left(\frac{\rho_{dm}^{MW}f}{M}\right)\pi R_{wd}^2\frac{v_{esc}^2}{v}\sim 2.7*10^{20}f\left(\frac{M}{g}\right)^{-1}\left(\frac{R_{wd}}{5600 \text{km}}\right)^2\left(\frac{v_{esc}}{2.3*10^{-2}}\right)^2 \text{Gyr}^{-1},
\end{equation}
where $R_{wd}$ and $v_{esc}$ are the WD's radius and escape velocity, respectively.

Once inside, EMBHs must sink to the WD core in order to pair and annihilate.  The EMBH mass and WD the core density and magnetic field  determine an EMBH's behavior.  The core magnetic field separates lighter EMBHs of opposite charge, preventing them from annihilating initially. Heavier EMBHs can overcome the magnetic field, and friction with the WD directs their behavior instead.

To annihilate, light EMBHs must first sink to where the magnetic and gravitational forces from the WD balance the attractive force toward other EMBHs.
The strength, prevalence, and structure of WD magnetic fields remain an area of active research.   The surface magnetic field of  WD J0551+4135 is $\sim 50$ kG \cite{Hollands_2020}.   Current measurement techniques only detect surface magnetic fields and are only sensitive to those $\gtrsim $1 kG \cite{2018A&A...618A.113B}.  When considering $1 M_{\odot}$ WDs, we only use those with magnetic fields too small to measure ($\lesssim 1$ kG).  We argue that the core magnetic field is equal to the observed surface field.  This is justified because magnetic fields in WDs are thought to either be fossil fields left over from the progenitor star or the result of differential rotation between proto-WDs in a binary system and the common envelope they formed in  \cite{2018CoSka..48..228K}.  In the former case, the WD magnetic field arises from conserving the magnetic flux of the larger progenitor.   Modeling indicates that the poloidal fossil field can only persist over long time scales  if it is supported by a twisted toroidal field outside of the core   \cite{Braithwaite_2004}. In the latter case, the magnetic field comes from a magnetized accretion disk, which settles onto the surface of the WD  \cite{Nordhaus_2011}.  In both cases, the magnetic field is generated well outside of the core, leaving no reason to expect the core magnetic field to be much larger than that at the surface. 

A sufficiently strong poloidal magnetic field will separate oppositely charged EMBHs and prevent them from interacting.  Interactions between EMBHs of the same sign charge are comparatively weak and neglected in this estimate.  One can estimate where EMBHs will settle by balancing the attractive forces of gravity and the opposite-sign EMBHs with the separating force of the magnetic field: 
\begin{equation}
\label{simplebalance}
    -B_{int}\frac{M}{m_{P}} +\frac{4\pi r}{3}\frac{\rho_{c}}{m_{P}^2}M+ 2\frac{M^2(N-1)}{m_{P}^2 (2r)^2} =0.
\end{equation}
Here, $B_{int}$ is the magnetic field at the center, $\rho_{c}$ is the local density, $r$ is the radial distance from the core  and $N$ is total number of EMBHs in the core, which we assume are split evenly by charge.  Note that the attraction between EMBHs  and the opposite charge cluster is doubled due to the combined magnetic and gravitational attraction.  If only one EMBH is present in the star, it will settle at $r = \frac{3}{4\pi}\frac{B_{int}m_{P}}{\rho_{c}}$, which for a 1 $M_{\odot}$ WD with a core density of $\sim3* 10^7$ g/cm$^3$ and a core magnetic field strength of 1 kG is $r\sim 0.03$ cm. We determine the time needed to reach this region using Eq.\eqref{WDfric} and a density profile derived in  Ref.\cite{Fedderke_2020} and find all EMBHs light enough to be separated by the core magnetic field sink to their equilibrium point in well under 1 Gyr.

As N increases, the stable solution to Eq.\eqref{simplebalance} moves closer to the center and then disappears.  At this point, EMBHs fall to the center of the WD instead of separating into polarized regions.  This happens when 
\begin{equation}
    B_{int}< \text{Min}\left[\frac{4\pi r}{3}\frac{\rho_{c}}{m_{P}}+\frac{1}{4}\frac{M(N-1)}{m_{P} r^2}\right],
\end{equation}
 which is met when $N$ satisfies
\begin{equation}
\label{massneeded}
    \frac{B_{int}^3m_{P}^3}{3\pi^2\rho_{c}^2}=2.2\times10^2g \left(\frac{B_{int}}{10^3G}\right)^3\left(\frac{10^8 \text{g cm}^{-3}}{\rho_{c}}\right)^2<M(N-1).
\end{equation}
Here `Min' refers to the minimum as a function of $r$.
EMBHs in separated clusters moving at thermal velocity do not have enough kinetic energy to overcome the magnetic field and reach each other until Eq.\eqref{massneeded} is satisfied.  When it is satisfied, mutual attraction between the EMBH clusters dominate over other forces, and annihilation proceeds quickly.  Only a single annihilation is needed to destroy the WD.  

EMBHs heavy enough to satisfy Eq.\eqref{massneeded} when $N=1$ are not affected by the core magnetic field.  They fall toward the core and interact with opposite sign EMBHs when their mutual attraction becomes stronger than their gravitational attraction to the WD core.  This happens once they fall within a radius $r_* \simeq \left(\frac{3M}{8\pi \rho_{c}}\right)^{1/3}\sim9$ cm $\left(\frac{M}{10^{12}g}\right)^{1/3}\left(\frac{10^8 \text{g cm}^{-3}}{\rho_{c}}\right)^{1/3}$.  For both the 1 $M_{\odot}$ and 1.14 $M_{\odot}$ WD, the time EMBHs need to sink to $r_*$ scales approximately linearly with their mass, with $4\times10^{12}$ g being the heaviest EMBHs that can reach the core  within a Gyr.  The large friction with the WD should dissipate the angular momentum and kinetic energy of infalling EMBHs, leading them to sink directly to $r_*$. They then only need to traverse $r_*$ to meet and annihilate.  We can conservatively estimate the time to reach each other as $r_*/v_*$, where $v_*$ is the EMBH velocity at the core, which scales roughly as $v_*\sim 10^{-21} (M/10^{12}\text{g})^{-1}$ in the 1.14 $M_{\odot}$ WD case and faster in the 1 $M_{\odot}$ WD case due to their lower core density.  Even the heaviest, slowest EMBHs that make it to the core in under a Gyr annihilate in well under a Gyr.

When two opposite charge EMBHs collide, they release energy through a magnetic ``chirp'' during the collision and through Hawking radiation of the non-extremal black hole left behind afterward.  The magnetic dipole they represent vanishes in about $t_s\sim 2r_s=4GM\sim10^{-38}\left(\frac{M}{\text{g}}\right)$ s, the time it would take an EM signal to traverse the Schwarzschild radii, $r_s=2GM$, of the two EMBHs.  One can expect an EM ``chirp'' from the sudden changing and rearranging of the magnetic field, akin to the gravitational chirp that happens during the final moments of an uncharged binary merger and the subsequent ringdown.  A magnetic dipole, $m$, will radiate energy at a rate $\frac{dE}{dt} = \frac{2}{3}\left(\frac{d^2m}{dt^2}\right)^2$.  Just before colliding, two EMBHs separated by a distance $2r_s$  will form a dipole with a dipole moment $2r_s*Q \sim 4\frac{M^2}{m_{P}^3}$.  If this were to suddenly disappear in $t_s$, a burst of $\frac{2}{3}M\sim 3.7\times10^{23}\left(\frac{M}{\text{g}}\right)$GeV  of EM energy would be emitted.  The energy needed to destroy each of the WDs is shown in Table \ref{tab:table1}.  The ``chirp'' alone allows any EMBH with $M\gtrsim 0.02$ g to furnish the energy needed to destroy a 1$M_{\odot}$ WD and with $M\gtrsim 8\times10^{-5}$ g to supply the energy needed to destroy WD J0551+4135.

EMBHs can also supply enough energy to destroy a WD through the Hawking radiation emitted by the non-extremal remnant left behind by the merger. Ref.\cite{Fedderke_2020}  suggests a black hole of mass $M$ inside of a WD will Hawking radiate at a rate of $\sim1.4\times10^{49}$ GeV s$^{-1} \left(\frac{M}{g}\right)^{-2}$.  If we assume the EMBHs involved in the merger have approximately equal masses, then the remnant left behind would then be non-extremal and need to lose $\sim M$ of mass to become extremal again.  These radiate at nearly the same rate as uncharged black holes, and may radiate even faster if we consider Q enhancements to the decay rate described in Refs.\cite{Maldacena:2020skw,bai2020phenomenology}.  Assuming the merger remnant radiates at least $\sim1.4\times10^{49}$ GeV s$^{-1} \left(\frac{M}{g}\right)^{-2}$, EMBHs with masses as low as 0.012 g for 1 $M_{\odot}$ WDs and $2\times10^{-5}$ g for WD J0551+4135 will be able to supply the energy needed to destroy their host WD.  Works such as Ref.\cite{Dvali:2020wft} suggest that the semi-classical description of Hawking emission is not correct and that black holes may radiate different amounts of energy, with different characteristics and over different time scales than predicted above.  If true, this changes the mass range of EMBHs able to destroy a white dwarf.  If the emission rate slows down, then the heaviest EMBHs may not radiate quickly enough to initiate runway fusion. If the amount of radiation is reduced, then the lightest EMBHs may not emit enough energy to trigger fusion.  Precisely how this mass range would evolve depends on the details of how Hawking emission changes outside of the semi-classical approximation, which remains uncertain \cite{Dvali:2020wft}.  This does not  particularly affect the bounds here, as most EMBHs able destroy a WD through Hawking radiation can also do so through the magnetic chirp produced when they annihilate.

We limit the abundance of EMBHs so that WDs do not accumulate enough to overcome the magnetic field barrier and annihilate within 1 Gyr (see  Eq.\eqref{massneeded}).  The bound for EMBHs too heavy to be affected by the WD core magnetic field is set by requiring that WDs not accumulate 6 EMBHs in 1 Gyr.  In a group of 6 EMBHs there is a $>95\%$ chance that at least one has the opposite charge as the others and is able to annihilate.  The existence of hundreds of non-magnetic WDs with masses around 1 M$_{\odot}$ and cooling ages longer than 1 Gyr, without any apparent suppression based on age or mass  \cite{2017ASPC..509....3D}, rules out EMBH parameter space where annihilation within 1 Gyr would be expected.  For the 1 $M_{\odot}$ WDs this bound is 
\begin{equation}
f<
    \begin{cases}
    1\times 10^{-17}& 0.012\text{ g}\leq M\leq 4.3\times 10^2 \text{ g}\\
    2.3\times 10^{-20}\left(\frac{M}{g}\right) & 4.3\times 10^2 \text{ g}<M<4\times 10^{12}\text{g}
    \end{cases},
\end{equation}
and for WD J0551+4135 this is
\begin{equation}
f<
    \begin{cases}
    2\times 10^{-13}& M\leq 8.3\times 10^6 \text{ g}\\
    2.4\times 10^{-20}\left(\frac{M}{g}\right) & 8.3\times 10^6 \text{ g}<M<4\times 10^{12}\text{g}
    \end{cases}.
\end{equation}
\subsection{Binary Mergers and Gamma-Ray Emission}
\label{sec:binarymerger}

Oppositely charged EMBHs can form binary systems in the early Universe.  These binaries spin down and eventually merge.  Assuming the EMBHs involved have approximately equal and opposite charges, the merger leaves behind a new non-extremal black hole.  The new black hole will have lost a ``chirp mass'' fraction of its mass to EM and gravitational radiation and then, unprotected by an extremal charge, would rapidly Hawking radiate down to its new extremal mass.  Conservatively, we assume it will radiate $\sim M$, the mass of the EMBHs.  The hot radiation from these merger remnants gets reprocessed by the thermal bath, and reaches Earth as a gamma-ray signal.  We can place constraints on the EMBH abundance by requiring this to be less than the 2$\sigma$ upper limit on the diffuse extra-galactic gamma-ray flux reported in the most recent isotropic background analysis  of data collected by the Fermi-LAT Telescope  \cite{Ackermann:2014usa}.

A binary decouples from the Hubble flow and begins spiraling inward when two opposite charge EMBHs' free-fall velocity toward each other exceeds the Hubble velocity pulling them apart.  The unique redshift, $z_d$, when this happens is determined by $M$ and by the co-moving separation between the EMBHs, $x$.  The time for a binary to merge, $t_m$, is determined by $x$, $z_d$, and the distance $y$ to the next nearest EMBH. EMBHs do not interact strongly with others of the same sign charge, so only one EMBH in the binary will interact with the next nearest charge.  The next nearest EMBH will tug the in-falling opposite sign one out of its free-fall path, causing the newly decoupled system to form an eccentric binary instead of immediately annihilating in a head-on collision.  The closer the next nearest EMBH, the smaller the eccentricity of the binary system.  Less eccentric binaries take longer to spin down.  This all affects the merger rate per black hole for a given redshift, $\Gamma_m(z)$, which is derived in full in Appendix \ref{sec: app merge}.

When a binary merges, it produces a sub-extremal MBH that begins to decay.  An uncharged black hole decays in 
  \cite{1974Natur.248...30H, Fedderke_2020}
\begin{equation}
    t_{bh}\sim\frac{15360\pi}{153/8}\frac{M^3}{m_{P}^4}\sim 1.3\times10^{-26}\left(\frac{M}{g}\right)^3  s,
\end{equation}
which is a reasonable estimate for the time a non-extremal MBH needs to decay down to it's near-extremal mass (this is, in fact, an over estimate, as we only included degrees of freedom up to MeV).   This is fast compared to the age of the Universe for all black holes constrained by radiation from this decay (roughly $\lesssim 10^7$ g), so it is reasonable to treat the entire decay as simultaneous with the merger.  Recent works   \cite{Maldacena:2020skw,bai2020phenomenology} suggest that the decay rate for non-extremal magnetic black holes is enhanced by a factor of the black hole charge, $Q$, This would reduce the decay time by $\frac{1}{Q}$, further justifying our decision to treat all decays as instantaneous.

Sub-extremal black holes in this mass range radiate both electromagnetically and hadronically. The hadronic radiation quickly fragments into more stable particles, including neutral pions.  These then decay into 2 photons.  A gamma-ray signal comes from both the photons radiated directly by the black hole and  those resulting from pion decays.  We estimate the combined photon spectrum $\frac{d^2N_{\gamma}}{dE_{\gamma}dt}$ using the derived analytic expressions for a decaying black hole's photon spectra found in equations (30)-(37) of   Ref.\cite{Ukwatta_2016}.    While the merger remnant's mass still exceeds its charge by an $\mathcal{ O}(1)$ factor it will radiate with approximately the same surface temperature as a non-magnetic black hole.  We estimate the total photon spectrum produced during this decay as $\frac{dN_{\gamma}}{dE_{\gamma}} \simeq t_{bh}\frac{d^2N_{\gamma}}{dE_{\gamma}dt} $,  as the black hole will radiate most of it's energy in this time and around its initial mass.  Ref.\cite{Dvali:2020wft} suggests that the semi-classical description of Hawking radiation may be inaccurate.  If true, one would estimate the EMBH bounds using a modified photon emission spectrum derived from the new description of black hole emission.  The bounds on EMBH abundance primarily depend on the total energy of radiated photons able to scatter with the thermal bath, those with $E_{th}=\frac{36000}{1+z}$GeV, where $z$ is the emission redshift. These bounds will scale inversely with the fraction of hawking radiation emitted above $E_{th}$. 

Photon-photon interactions with CMB photons and Inverse Compton scattering off electrons reprocess the initial photon energy distribution from the binary annihilation.  The reprocessed gamma-ray spectral energy distribution has the form   \cite{Berghaus:2018zso,1989ApJ...344..551Z}
\begin{equation}
    L(E_\gamma,z) =
\begin{cases}
0.767 E_{th}(z)^{-0.5}E_\gamma^{-1.5} & E_\gamma\leq0.04E_{th}(z)\\
0.292E_{th}(z)^{-0.2}E_\gamma^{-1.8} & 0.04 E_{th}(z)<E_\gamma<E_{th}(z),
\end{cases}
\end{equation}
where $E_{th}=\frac{36000}{1+z}$GeV is the threshold energy a photon must have to be reprocessed, $E_{\gamma}$ is the energy of the photon, and $z$ is the redshift at the time of emission.  Photons with initial energies below $E_{th}$ free stream to Earth  \cite{1989ApJ...344..551Z}, while those at higher energies are reprocessed.   A total reprocessed energy spectrum can be found by multiplying $L(E_{\gamma},z)$ by the amount of energy radiated above $E_{th}$.  The shape of the spectrum is independent of this energy.  The parts of the initial spectrum, $N_{Einit}(E_{\gamma},z)$, that were and were not reprocessed combine to give a total visible spectrum 

\begin{equation}
    N_{Etot}(E_{\gamma},z) =
    \begin{cases}
    N_{Einit}(E_{\gamma},z)\theta(E_{th}-E_{\gamma})+L(E_{\gamma},z)\int_{E_{th}}^{\infty} N_{Einit}(E_{\gamma}',z)_{init} E_\gamma' dE_{\gamma}' & E_{\gamma}\leq E_{th}(z)\\
    0 &E_{\gamma}> E_{th}(z).
    \end{cases}
\end{equation}
Here, $\theta(x)$ represents the Heavyside function.  The flux per unit energy of gamma-rays that reach Earth can be found by convoluting the reprocessed spectrum with the merger rate as a function of redshift:
\begin{equation}
\label{extragalflux}
    F_{\gamma}(E\gamma_0) = \frac{\rho_{dm} f}{4\pi}\int_0^{z_m}\frac{1}{H(z)} \Gamma_m(z)N_{Etot}\left(E_{\gamma0}(1+z),z\right) dz,
\end{equation}
where $H(z)$ is the Hubble rate at $z$, and $E_{\gamma0}$ is the photon energy today, and $\Gamma_m(z)$ is the merger rate per black hole at $z$. $z_m=\sqrt{\frac{E_{th}(z=0)}{E_{\gamma0}}}-1$ is the highest redshift at which a photon with observed energy $E_{\gamma0}$ could be emitted and still reach Earth.  Any photon  with $E_{\gamma0}$ that reaches Earth today must have been emitted at a redshift $z<z_m$ because those emitted at earlier times will be reprocessed and redshifted down to lower energies.  The total flux per Fermi energy bin is found by integrating $F_{\gamma}(E_{\gamma_0})$ over the energy range for that bin.  We require the integrated gamma-ray fluxes to be less than the 2$\sigma$ upper bound observed by Fermi. The strongest constraints come from the highest energy bin, 580 GeV-820 GeV.

We have, so far, neglected friction with the thermal bath.  As explained in Section \ref{warmgas}, EMBHs interact strongly with plasmas such as the thermal bath. Friction slows infalling EMBHs in binaries, causing them to decouple at later times and get stretched further apart by the Hubble flow.  The stretching reduces the number of binaries that form and causes those that do to merge at later times.  The result is that EMBHs that are still strongly interacting with the thermal bath at $z_d$, their friction-free decoupling redshift, do not significantly contribute to either the merger gamma-ray signal or to the EMBH abundance bound.  We consider an EMBH to be strongly interacting with the thermal bath when friction more strongly influences its velocity than Hubble acceleration, $\left(\frac{dE}{dx}\right)_f/(MV)>H(z)$.  The frictional force grows as $M^2$, making heavier EMBHs more sensitive to it.  Friction begins to weaken the bound for EMBHs with  $M\sim10^6$g and removes it entirely for those with $M>10^7$g.
\begin{figure}
    \centering
    \includegraphics[scale=0.5]{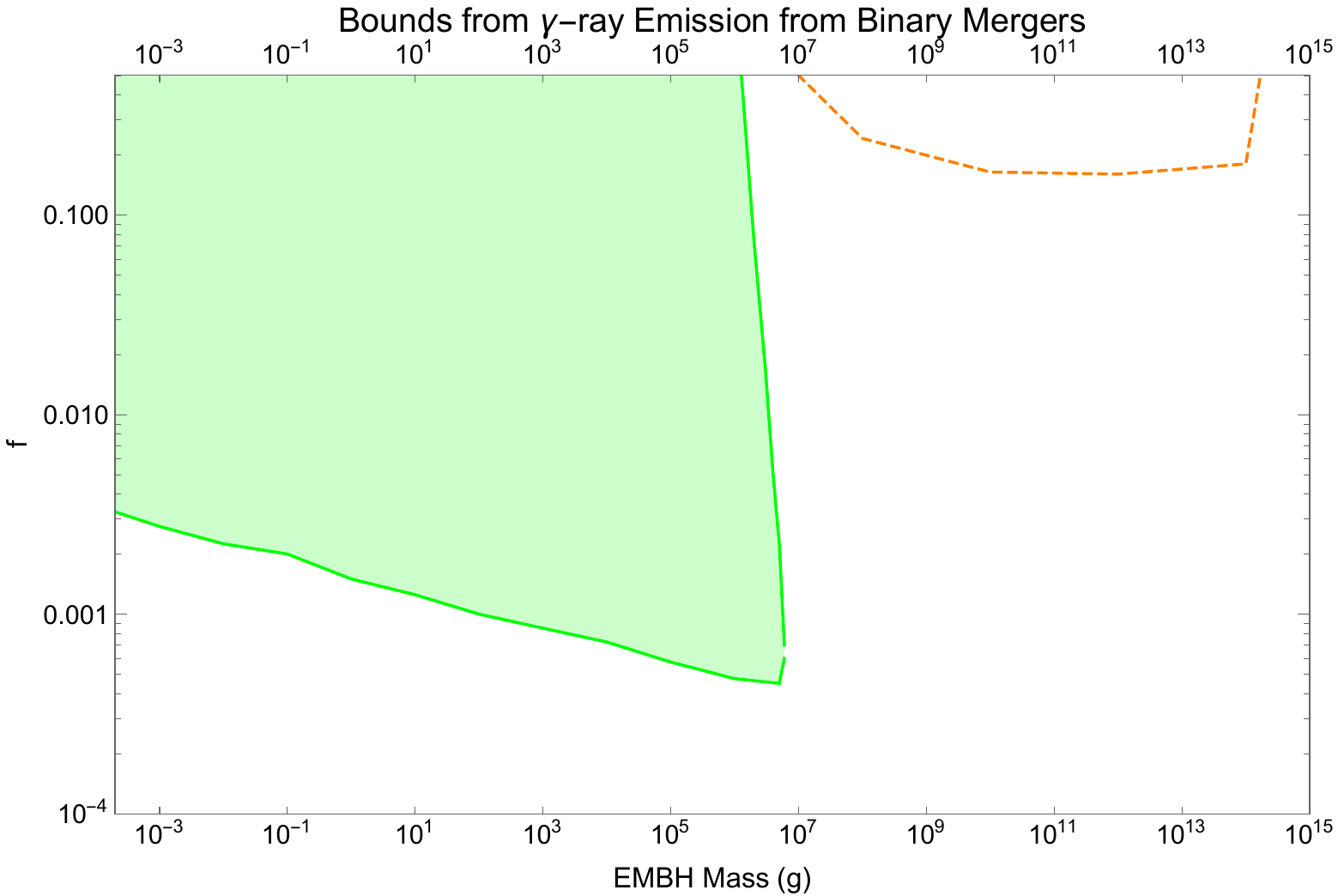}
    \caption{The green region shows the bounds on the EMBH abundance, $f$, based on gamma-ray emission from mergers of EMBH binaries not strongly influenced by friction with the thermal bath.  The dotted orange line gives a rough estimate of the bounds that arise from gamma-ray emission from binaries whose initial infall dynamics were strongly influenced by friction with the thermal bath. The friction-free bounds weaken for high $f$ for EMBHs with $10^6$ g $<M<$ $10^7$ g  because less eccentric binaries are more sensitive to frictional effects and raising $f$ lowers the typical eccentricity.}
    \label{fig:binarybound}
\end{figure}

The bounds are shown in Fig.\ref{fig:binarybound}.  The most notable feature is that they weaken at higher $f$ for EMBHs with masses $10^6-10^7$g.  Increasing $f$ decreases the typical spacing between EMBHs, which reduces the distance between a binary and the next nearest EMBH, lowering the eccentricity of the most eccentric binaries. The merger rate is dominated by these binaries, so decreasing the maximum eccentricity slightly lowers the merger rate.  This is normally insignificant, but becomes important when friction suppresses binary formation. 

Weaker bounds may be found for heavy EMBHs ($M>10^7$g) by accounting for friction.  These EMBHs form bound pairs if they can traverse the distance separating them within one Hubble time while infalling at a terminal velocity set by friction, their charge, and their initial separation.  Friction reduces the number of bound pairs, so the constraint from gamma-ray emission for EMBHs with $M>10^7$g is weak compared to the other new constraints presented in this paper.  For comparison, we show a rough constraint for EMBHs strongly affected by friction in Fig.\ref{fig:binarybound}, but do not include this in our final set of constraints.

Ref.\cite{Maldacena:2020skw} suggests that fermionic radiation from near extremal MBHs is be enhanced by a factor of $Q$, reducing the black hole lifetime by a factor of $\frac{1}{Q}$, without changing the surface temperature.  This should not significantly change the the photon spectrum that reaches Earth from mergers of EMBHs with $M<10^7$g. The low energy portion of the spectrum comes from hadronic decays. If the surface temperature remains unchanged, the hadronic spectrum should have the same form, so the low energy part of the spectrum is unchanged.   Reprocessed photons contribute to the high energy part of spectrum.  The $Q$ enhancement may suppress the number of direct photons produced due to the reduction in the black hole lifetime.   However, electrons and positrons would be produced at a $Q$ enhanced rate, and produce the same spectrum as photons after reprocessing by the thermal bath \cite{1989ApJ...344..551Z}.

\subsection{Destruction of Large Scale Magnetic Fields}
\label{sec:destruction of Mfield}

EMBHs interact with and can destroy large scale coherent magnetic fields.  The bounds from this are subdominant to those derived in this paper, so we only touch on them briefly.

\textbf{Parker Bound}
There are multiple versions of the Parker bound  \cite{Lewis:1999zm, PhysRevLett.70.2511, Turner:1982ag} on the abundance of magnetic monopoles based on their interaction with the Milky Way's magnetic field.  This field accelerates magnetic charges, and drains its energy in the process.  The magnetic charges must not destroy the magnetic field or prevent it from forming.  The protogalactic Parker bound offers the most stringent monopole abundance constraints  \cite{Lewis:1999zm}.  It limits the magnetic charge density present during the collapse of the protogalaxy, when the Galactic magnetic field was only a small seed field unsupported by a Galactic dynamo.  When applied to EMBHs in the Milky Way this gives
\begin{equation}
    f<0.06.
\end{equation}
Stronger constraints, $f<4\times 10^{-4}$ \cite{bai2020phenomenology} and $f<1.7\times 10^{-3}$ \cite{ghosh2020astrophysical}, were found by applying a Parker type bound to the Andromeda Galaxy whose magnetic field is coherent over larger length scales than those of the Milky Way.  Uncertainty in this bound comes from uncertainty in the properties of the magnetic field in Andromeda \cite{bai2020phenomenology}.  We show the bound from Ref.\cite{bai2020phenomenology} in Fig.\ref{fig:catminlim} for comparison.

\textbf{Cluster Magnetic Fields}
The largest scale magnetic fields  observed in galaxy clusters are in radio relics.  These are coherent over $\sim 2$ Mpc  \cite{Bonafede:2013ewa, Kierdorf_2017}.  EMBHs in these clusters would form a diffuse magnetic plasma, suppressing magnetic fields that extend over scales larger than its Debye length, $\lambda_d$.  Requiring $\lambda_d >2$ Mpc for an EMBH plasma in a typical galaxy cluster gives the bound
\begin{equation}
    f<0.7.
\end{equation}

\subsection{MACRO}
\label{sec:MACROnocat}
MACRO was a dedicated monopole detector that ran from 1989 through 2000 and relied on ionization effects to detect passing magnetic charges \cite{Ambrosio:2002qq}.  While the detectors and analyses focused on Dirac monopoles with $Q=g_D$  \cite{Ambrosio:2002qq}, they were at least as sensitive to higher charge magnetic particles  \cite{Derkaoui:1998gf}, and have been used to constrain a variety of similar objects, including extremal electrically charged black holes \cite{Lehmann:2019zgt}.  MACRO flux bounds should therefore be applicable to EMBHs.  Assuming EMBHs do not cause nucleon decays, to be discussed in Section \ref{sec:BoundsCat}, MACRO places an upper limit on the EMBH flux of \cite{Ambrosio:2002qq}
\begin{equation}
    F<2.8\times10^{-16} \text{cm}^{-2}\text{sr}^{-1}\text{s}^{-1}.
\end{equation}
This corresponds to a bound
\begin{equation}
    f<217\left(\frac{M}{\text{g}}\right).
\end{equation}
\section{The Effects of Proton Decay Catalysis on EMBH Bounds}
\label{sec:BoundsCat}

As discussed in the introduction, EMBHs might catalyze proton decay, perhaps analogous to the effect in monopoles from grand unified theories  \cite{Callan:1983nx,Rubakov:1982fp}.  However, uncertainty around what lies at the `core' of the EMBH (the strength of B-violation, the existence of short-distance hair etc.)  makes the cross section for this process model dependent. 

We circumvent many of these complications by looking for robust bounds independent of the cross section for catalysis.  Inside of WDs, heat produced by catalysis can generate convection currents that prevent the EMBHs from annihilating, weakening the bounds from WD destruction.  On the other hand, catalysis can cause anomalous heating of NSs causing x-ray emission.  In addition, excess heating of WDs can set a (weaker) bound, while catalysis in the Sun can be constrained by neutrino detectors, such as Super Kamiokande.  EMBH abundance constraints from the MACRO detector are weakened but still present when accounting for catalysis effects. Whatever the catalysis cross section, we can place a robust bound on the EMBH abundance at any mass.  

We estimate the most robust bound by finding the threshold catalysis cross section needed to activate convection in WDs, $\sigma_{th}$.   This minimal cross section is then used to find bounds from WD heating.  Increasing the cross section above $\sigma_{th}$ increases the heating in WDs, strengthening the heating bound.  Reducing the cross section below $\sigma_{th}$ will allow EMBH annihilations to resume in WDs.    For a set cross section, NS heating generates stronger bounds than WD heating.  However, many estimates of the catalysis cross section depend on the velocity of passing nuclei  \cite{PhysRevLett.50.1901, Bracci:1983fe, Olaussen:1983bm}, which are quite different in WD ($\sim 10^{-4}$) and NS ($\sim 0.3$) cores, making it unreasonable to assume EMBHs will have the same effective cross sections in both environments.  
We only compare bounds that rely on catalysis in WDs, where we can give a more consistent description of the cross section.  Catalysis bounds in the Sun and in NSs are described and shown for a specific cross section for comparison and future application to specific models of EMBHs.  

Whether $\sigma_{th}$ is physically reasonable depends on two different length scales set by the EMBH and by the properties of the WD core: a geometric scale and a scattering length.  One might expect a catalysis cross section at the geometric scale where the energy scale of the magnetic field is larger than the proton mass and the dynamics are better described by two-dimensional degenerate radial fermions  \cite{Maldacena:2020skw}. This cross section will be referred to as the QCD cross section.  
\begin{equation}
\label{QCDcross}
      \sigma_{QCD} = \sqrt{\frac{\pi}{8}} \frac{M}{m_{P}\Lambda_{QCD}^2}\sim 2\times 10^{-22}\text{cm}^2\left(\frac{M}{g}\right),
\end{equation}
where $\Lambda_{QCD} \equiv 218$ MeV is the approximate QCD scale.  All nuclei entering this region will have dynamics dominated by the background magnetic field and could plausibly have non-negligible overlap with a proton-decaying core.  We note that this cross section can be altered by an angular momentum barrier between the strong magnetic field of an EMBH and the electric charge and magnetic moment of an incoming nuclei \cite{PhysRevLett.50.1901}.  

While $\sigma_{QCD}$ defines the region where nucleon decay may occur, the strong magnetic field around EMBHs allows them to have even larger effective catalysis cross sections.    For example, charged nuclei with non-zero magnetic moments accelerate and radiate photons when passing through the magnetic field near an EMBH, causing them to fall into a bound state which overlaps $\sigma_{QCD}$ \cite{Bracci:1983fe,Olaussen:1983bm}.   Any enhancement to the effective catalysis cross section is limited by scattering in the WD medium.  Nuclei falling toward the EMBH can scatter and regain enough kinetic energy to escape before reaching the inner decay region. This introduces the second important length scale: the  scattering length of the nuclei in the local medium,
\begin{equation}
\label{scatwd}
    l_s = \frac{m_n}{\rho_{c}}\frac{1}{\pi \lambda_d^2}\sim5\times10^{-8}{\rm \left(\frac{T}{10^6 \text{K}}\right)^{-1}\, cm}
\end{equation}
Here, $m_n$ is the mass of the nucleus involved, and $\lambda_d$, $T$, and $\rho_c$ are the Debye length, temperature, and density in the WD core, respectively. 

The length scales described above set effective upper limits on $\sigma_{th}$.  Given that $\sigma_{QCD}$ describes a region where nucleon decay may occur, any value of $\sigma_{th}\leq\sigma_{QCD}$ is physically plausible.  
Mechanisms exist to allow $\sigma_{th}$ to grow larger than $\sigma_{QCD}$  \cite{Bracci:1983fe,Olaussen:1983bm}, but this is limited by nuclear scattering.  
$\sigma_{th}$ is only physically plausible if
\begin{equation}
\label{eq:Sigmax}
    \sigma_{th}\leq \sigma_{M}\equiv \text{Max}[\pi l_s^2, \sigma_{QCD}],
\end{equation}
where we define $\sigma_{M}$ as the maximal allowed catalysis cross section within the WD core.

\begin{figure}
    \centering
    \includegraphics[scale=0.4]{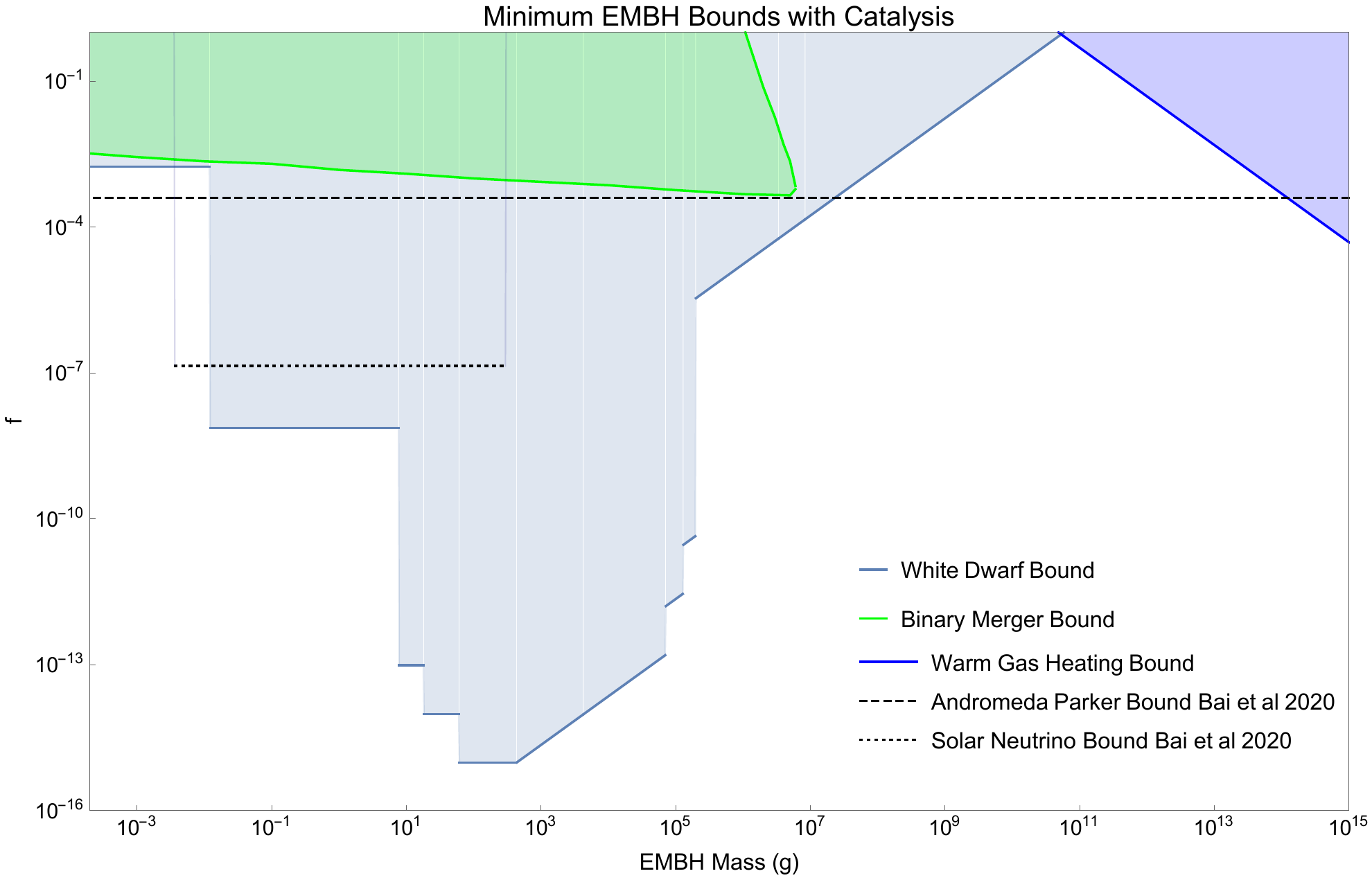}
    \caption{ The colored regions denote the minimum bounds on the EMBH abundance when catalysis effects are considered.  The WD overheating and annihilation bounds are related by catalysis effects and marked as one in light blue.  The bounds from overheating or destroying 1 $M_{\odot}$ WDs dominate for EMBHs with $M >0.012$g.   Lighter EMBHs cannot destroy a 1 $M_{\odot}$ WD when they annihilate so overheating WD J0551+4135 is used to constrain those with $M<0.012$ g.  No physically reasonable catalysis cross section could prevent EMBHs with $7\text{ g  }\lesssim M\lesssim 2\times10^5\text{ g }$ from annihilating in 1 $M_{\odot}$ WDs at all times, so the original annihilation bound remains in a somewhat weakened form.  The bounds appear jagged because the catalysis effects were considered at discrete cooling temperatures and would vary smoothly if these effects were considered at arbitrary cooling temperatures and times.  There are plausible catalysis cross sections that prevent EMBH annihilation outside of this mass range, and so bounds come from WD heating instead.   Bounds from gas heating and binary mergers are unaffected by catalysis and are described in Sections \ref{warmgas} and \ref{sec:binarymerger}, respectively.  Constraints derived in Ref.\cite{bai2020phenomenology} are shown in black for comparison and labeled Bai et al 2020.  The dashed line comes from a Parker bound applied to the Andromeda Galaxy (see Section \ref{sec:destruction of Mfield}).  The dotted line comes from energetic neutrinos produced from EMBH annihilations in the Sun (see Section \ref{sec:inSun}).  We cut this bound off for EMBHs with $M > 300$ g because convection in the Sun suppresses their annihilation rate.}
    \label{fig:catminlim}
\end{figure}

How we determine $\sigma_{th}$, the EMBH annihilation rate, and the WD heating bounds is elaborated on in the sections below.  Fig.\ref{fig:catminlim} shows the minimal bounds once catalysis effects are considered, along with the bounds not impacted by catalysis for a full picture of the constraints if catalysis is non-negligible.  We have made the conservative simplifying assumption that once convection starts in a WD, the EMBHs will be spread throughout the star and unable to annihilate.   In truth, they may still interact and annihilate under these conditions; however,  calculating how the annihilation rate and WD convective currents evolve with the catalysis cross section requires detailed modeling of convection in WD cores, which is beyond the scope of this paper and will only yield more stringent bounds.  

\begin{figure}
    \centering
    \includegraphics[scale=0.5]{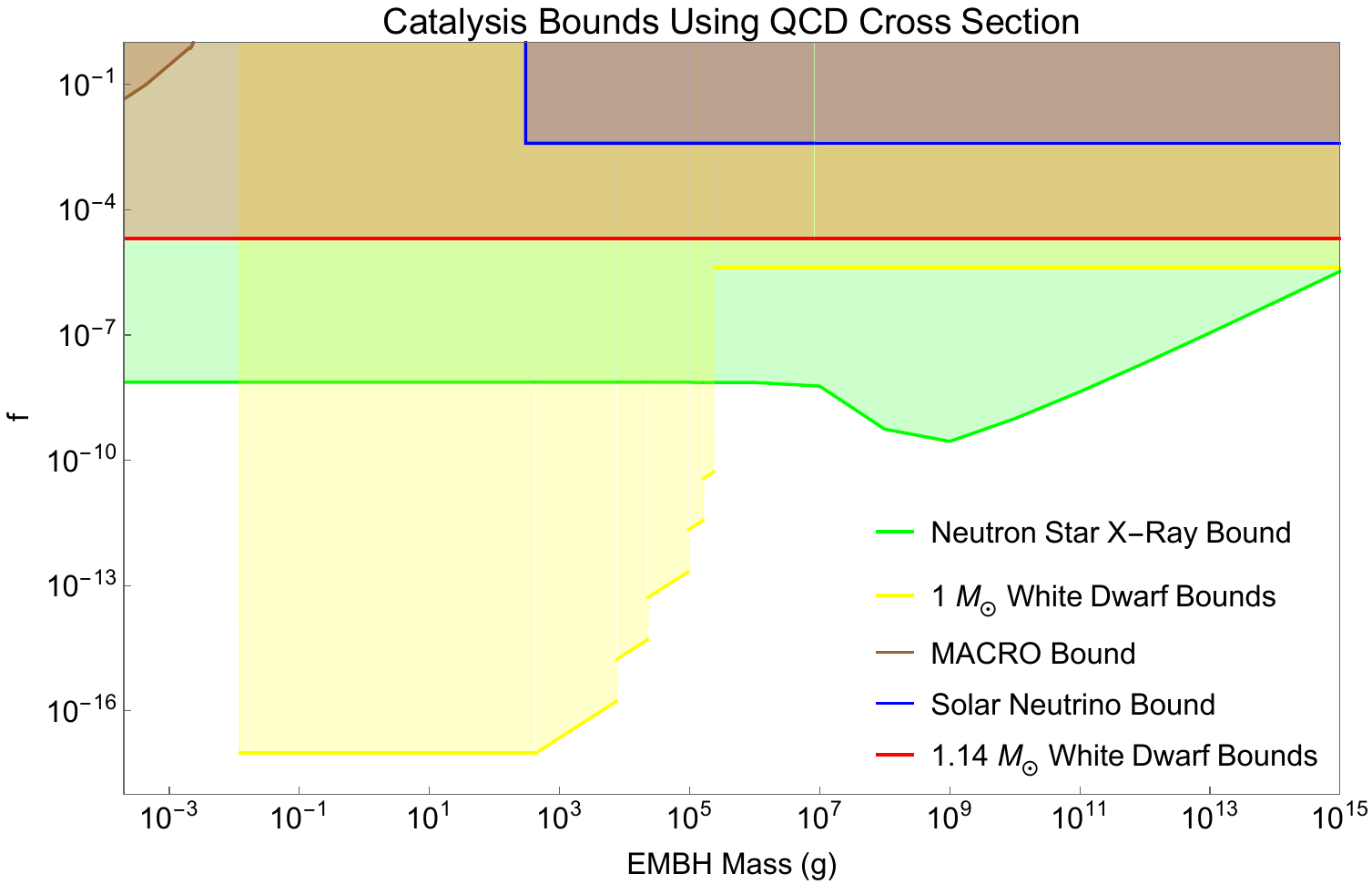}
    \caption{This plot shows catalysis bounds on EMBH abundance assuming $\sigma_{\text{cat}}=\sigma_{QCD}$.  The yellow and red lines denote bounds from 1 $M_{\odot}$ WDs and WD J0551+4135  respectively.  The annihilation bounds for 1 $M_{\odot}$ WDs  hold, though perhaps in weakened form for, EMBHs with $ M \lesssim 9\times10^5$ g.  These bounds appear jagged because the annihilation rate was estimated at discrete WD cooling times and would appear smooth if arbitrary cooling times were used instead. Heavier EMBHs initiate convection at all times and are constrained by overheating old WDs (Section \ref{sec:wdheating}). All EMBHs produce enough energy to start convection in WD J0551+4135 at all times, so only heating bounds are displayed.  The green region is ruled out by overheating old NSs.   This bound weakens for EMBHs with $M<10^7$ g because they annihilate inside of NSs, reducing the number that contribute to heating.   The dip in this bound appears because only a few EMBHs with $M>10^7$ g  are needed to overheat a NS, making annihilation unimportant (see Section \ref{sec:nsheating}).  The brown line marks constraints from a MACRO monopole search (Section \ref{sec:MACROcat}).  The blue line marks the limit from observations of energetic neutrinos produced by EMBH induced proton decay in the Sun.  This cuts off for EMBHs which sink to the stellar core and annihilate quickly ($M<300$ g). }
    \label{fig:catQCD}
\end{figure}

For illustration only, we display all of the bounds that arise if we assume the catalysis cross section is fixed at $\sigma_{QCD}$ in all environments in Fig.\ref{fig:catQCD}.  Here, additional bounds arise from physical processes, such as NS heating and neutrino production from proton decay in the Sun.

\subsection{Convection in White Dwarfs}
\label{sec:convect}

As explained in Section \ref{sec:wdann}, EMBH abundances are strongly constrained by their annihilation destroying WDs.  Catalysis heating bounds are significantly weaker but still relevant.  When an EMBH falls into a WD, or any medium, it can cause nearby nuclei  to decay into  lighter particles, releasing the nucleon mass energy and making the EMBH a source of radiation.  The energy radiated can create a severe enough temperature gradient to initiate convection in a WD core \cite{Freese:1984fz}.  Friction between the EMBHs and convecting material drags them away from center, potentially preventing them from annihilating within 1 Gyr and undermining the WD annihlation bound described in Section \ref{sec:wdann}.  The catalysis cross section determines the amount of heat generated in the WD core, which sets the temperature gradient, which itself determines whether convection occurs.  If convection does not happen, then the bound determined in Section \ref{sec:wdann} stands with some modifications.  Once convection starts, we assume the EMBHs, now spread  throughout the WD, do not annihilate.  Estimating the annihilation rate once convection begins requires detailed modeling of how EMBHs move within WD convection currents and would only yield stronger bounds. 

We determined whether EMBH annihilations could happen by comparing the minimum luminosity from EMBH induced nuclei decays needed to initiate  convection, $L_{th}$, to $L_{m}$, the maximum luminosity that can be radiated by EMBHs in the WD core assuming realistic catalysis cross sections and that enough EMBHs are present to annihilate under no catalysis conditions.  We determined $L_{th}$ for a variety of WD ages because younger WDs are hotter, and can accommodate more EMBHs before convection turns on.  This was done separately for both sets of white dwarfs considered.

Throughout the WD's conductive region, the temperature gradient is
\begin{equation}
\label{dTdr}
    \left(\frac{dT}{dr}\right)_{cond} = \frac{-L_{tot}}{4\pi \kappa r^2},
\end{equation}
where $L_{tot}$ is the combined luminosity from all of the EMBHs in the WD, $\kappa $ is its conductivity, and $r$ is the distance from the center of the WD.  We use the expression for $\kappa$ presented in    Ref.\cite{Potekhin:1999yv} for a WD made of equal parts carbon and oxygen with the density profiles for  1.14 $M_{\odot}$ and 1 $M_{\odot}$ WDs derived in Ref.\cite{Fedderke_2020}.     Two conditions must be met for convection to occur. First, there must be a region in the star which satisfies
\begin{equation}
\label{eqn:first condition}
   \left(\frac{dT}{dr}\right)_{ad}<\left(\frac{dT}{dr}\right)_{cond},
\end{equation}
where $T$ is the temperature,
\begin{equation}
    \left(\frac{dT}{dr}\right)_{ad} = \frac{1}{4}\frac{T}{P}\frac{dP}{dr}
\end{equation}
 is the adiabatic temperature gradient,
\begin{equation}
    P= \int_r^{R_{wd}} \frac{dP}{dr'}dr'
\end{equation}is the pressure, and 
\begin{equation}
    \frac{dP}{dr} = \frac{GM_{int}(r)\rho(r)}{r^2}
\end{equation} is the pressure gradient.   $M_{int}(r)$ is the mass contained within the radius $r$, and $\rho(r)$ is the mass density at $r$.  Starting with a surface temperature, $T_{s}$, we calculate $T$ and $\frac{dT}{dr}$ as a function of $r$ in km steps moving inward from the surface and in m steps for the inner-most kilometer of the WD.  At each step we check if the convective condition has been met.  WD interiors cool over time, causing older ones to develop more extreme temperature gradients and leading convection to start at lower decay luminosities than in younger stars.  Some EMBHs will start convection in cool Gyr-old WDs but not hotter 100 Myr-old ones.  To account for this, we considered the temperature gradient for WDs at cooling ages $\{10^9, 10^8, 10^7, 10^6, 10^5\}$ years and their corresponding internal temperatures $T_s=\{10^{6.8},10^{7.3},10^{7.7},10^{7.85}, 10^{7.95}\}$ K, as estimated in  Ref.\cite{DAntona:1990kji}.

If Eq.\eqref{eqn:first condition} is satisfied near the core, we check for the second condition needed to initiate convection.  Buoyant forces acting on the WD material must be greater than the viscous forces resisting deformation.  The Rayleigh number, $Ra$, parameterizes the relationship between these forces. 
For spherical geometries
\begin{equation}
    Ra = \frac{4\pi G\rho(r)D(r)}{3\chi \nu}r^5,
\end{equation}
where 
\begin{equation}
    D(r) = \frac{1}{T}\left[\left(\frac{dT}{dr}\right)_{cond}-\left(\frac{dT}{dr}\right)_{ad}\right]
\end{equation}
is the adiabatic excess at a given radius $r$  \cite{Freese:1984fz},
$\chi$ is the thermometric conductivity, and $\nu$ is the kinematic viscosity  \cite{2017mcp..book.....T}, which we take from   Ref.\cite{Woosley_2004}.  $Ra$ must be $\gtrsim10^3$ for convection to occur in a spherical system where heat is generated in the convection medium \cite{2002geod.book.....T}.

For each cooling temperature and set of WDs, we determined $L_{th}$, the minimum luminosity where  Eq.\eqref{eqn:first condition} and  $Ra\gtrsim 10^3$, are satisfied.  To determine if $L_{th}$ is physically plausible, we also calculated $L_{m}$,  the maximal luminosity that could be radiated from all of the accumulated EMBHs if they each had the largest possible cross section:
\begin{equation}
    L_{m} = N_{wd}\rho_c  \sigma_{M}v_n
\end{equation}
Here, $N_{wd}$ is the number of EMBHs present in the WD core, which we take to be the minimum number needed to destroy a WD if catalysis did not happen, set by Eq.\eqref{massneeded}, and $v_n$  is the thermal velocity.  Convection only starts if $L_{th}\leq L_m$.

\begin{figure}
    \centering
    \includegraphics[scale=0.6]{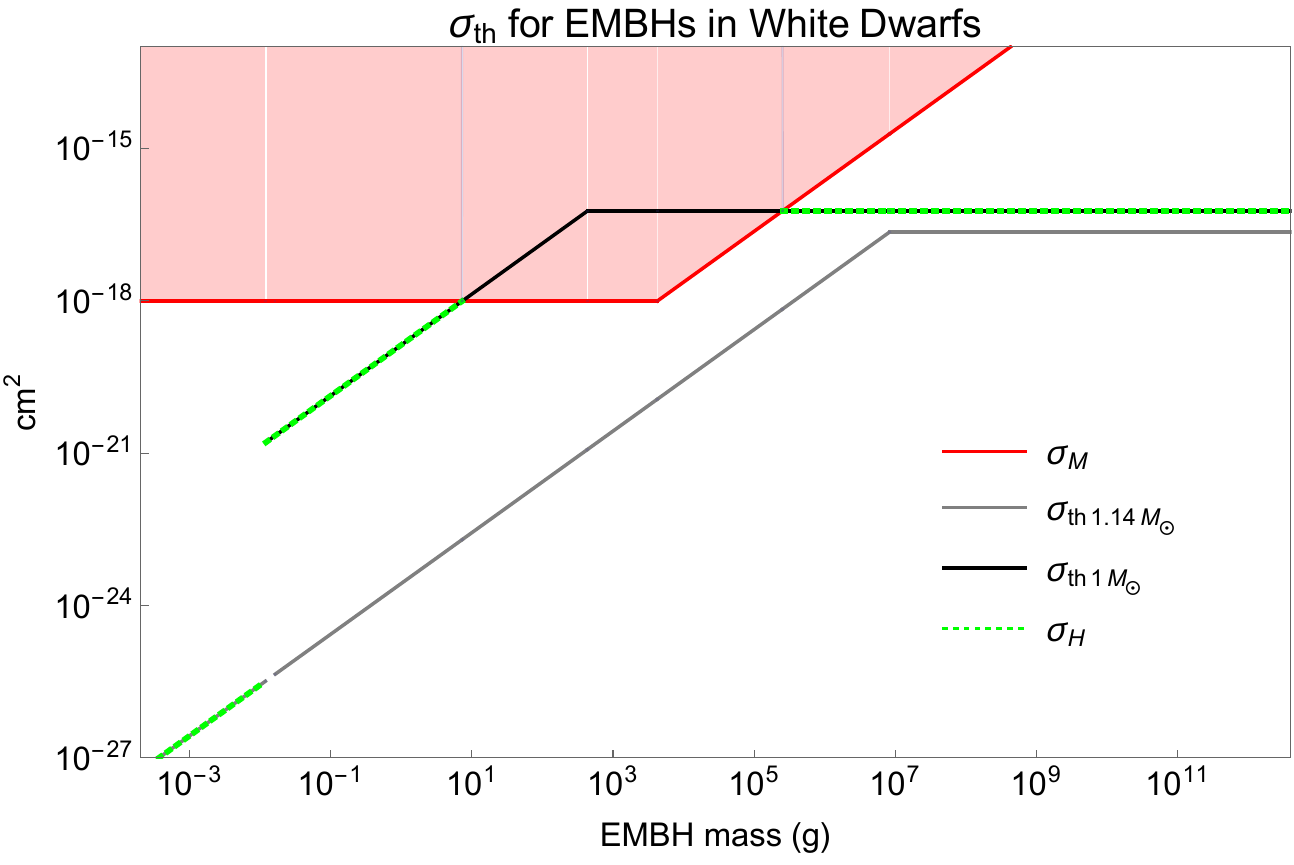}
    \caption{The black and gray lines above show the minimum catalysis cross sections, $\sigma_{th}$, needed to start convection at all cooling times considered in 1 $M_{\odot}$ WDs and WD J0551+4135, respectively.  The red region excludes implausible cross sections in the WDs, with the red line marking $\sigma_{M}$. No physically viable value for  $\sigma_{th}$ for 1 $M_{\odot}$ WDs exists for EMBH masses where the black line crosses into the red region. Here, some form of the original annihilation bound stands.  Heating bounds are set wherever the annihilation bounds break down ($M<8 \text{ g}$ and $2\times10^5$ g$<M$) using $\sigma_H$, which is marked by the green dotted line and follows whichever value of $\sigma_{th}$ for the two sets of WDs is larger. It follows $\sigma_{th 1.14 M_{\odot}}$ at low masses where no annihilation bound exists for  1 $M_{\odot}$ WDs.  $\sigma_M$ and $\sigma_{th}$ are temperature sensitive and are set for WDs at $T=10^{7.95}$ K which corresponds to a cooling time of $10^5$ years.  This is when the star is hottest and least sensitive to energy injections from catalysis. }
    \label{fig:multiplier}
\end{figure}

The original annihilation bounds remain unchanged if, at every temperature checked, $L_{th}> L_{m}$.  If $L_{th}> L_{m}$ at some, but not all, temperatures, then the annihilation bound is modified to prevent WDs from accumulating enough EMBHs for an annihilation to occur within the maximum time period for which $L_{th}<L_{m}$.  Finally, if $L_{th}\leq L_m$ at every temperature scale checked, then there exists a physically reasonable catalysis cross section that prevents annihilation at all times.  To find this, we first determine the catalysis cross section, $\sigma$ that corresponds to $L_{th}$ for each mass and cooling age:
\begin{equation}
    \sigma= \frac{L_{th}}{N_{acc}\rho_{c}v_n} .
\end{equation}
For a given mass, $\sigma_{th}$ is the maximum $\sigma$ found among all of the different cooling temperatures considered.  $\sigma_{th}$ is determined separately for 1 $M_{\odot}$ WDs and WD J0551+4135.   The maximum of the two is used to set the catalysis heating bounds in WDs and is referred to as $\sigma_H$. When $M<0.012$ g, the annihilation bound only applies to WD J0551+4135 and so only its value of $\sigma_{th}$ is used to set $\sigma_H$.  The calculated values of $\sigma_{th}$ and $\sigma_H$ can be found in Fig.\ref{fig:multiplier}.

\subsection{White Dwarf Heating}
\label{sec:wdheating}
The heat produced from EMBH catalyzed nucleon decays can make WDs older than a few billion years more luminous than expected. Observations confirm the existence of WDs with luminosities below $10^{-4}L_{\odot}$  \cite{1997ApJ...489L.157H,1988ApJ...332..891L} and ages above 8 Gyr  \cite{doi:10.1002/9783527636570.ch2, Kilic_2012,2017ASPC..509....3D}.   The EMBH abundance must be low enough to allow these old WDs to cool to their observed luminosities and  WD J0551+4135 to cool to its observed luminosity, $3\times 10^{-4}L_{\odot}$ \cite{Hollands_2020}, within its estimated lifetime of 1.3 Gyr \cite{Hollands_2020}.

The luminosity of EMBHs in a WD core is
\begin{equation}
    L=v_n \sigma_{H}\rho_c,
\end{equation}
with $v_{n}=\sqrt{\frac{3T_{wd}}{m_{n}}}$ and $\rho_c$ as the thermal velocity of nuclei and density in the core, respectively.  We take $T_{wd}$, the temperature of the WD core, to be  $10^{6.5}$K in a WD with $L =10^{-4}L_{\odot}$  and $10^{6.8}$K in WD J0551+4135, based on modeling in  Ref.\cite{DAntona:1990kji}.  The cross sections used for different EMBH masses, $\sigma_H$, are presented in Fig.\ref{fig:multiplier}. Some works \cite{PhysRevLett.50.1901,Bracci:1983fe,Olaussen:1983bm} suggest that the catalysis cross section can grow as the velocity of passing particles decreases.  If such a velocity dependence is present, then $\sigma_{H}$ for a young hot WD may correspond to a larger cross section in an older cooler one with lower thermal velocities.  We use $\sigma_{H}$ in our estimate because the velocity dependencies are model dependent, the thermal velocities of nuclei in the hottest and coolest WDs considered vary by less than an order of magnitude, and accounting for velocity dependencies will only strengthen the heating constraint.

A WD that has accumulated EMBHs for $t_{acc}$ years without annihilation will radiate with a combined luminosity
\begin{equation}
    L_{cat} = L\Gamma_{wd}t_{acc},
\end{equation}
where $\Gamma_{wd}$ can be found in Eq.\eqref{interaction rate}. One must keep $L_{cat}$ below $10^{-4}L_{\odot}$ for 1$M_{\odot}$ WDs older than $t_{acc}\sim 8 $ Gyr and  $3\times10^{-4}L_{\odot}$ for  WD J0551+4135, assuming $t_{acc}\sim$ 1.3 Gyr.  For WD J0551+4135 this age is a conservatively low estimate, as it is thought to result from the merger of two older WDs (1.3 Gyr ago) that would themselves have had more time to accumulate EMBHs \cite{Hollands_2020}.
The bounds on EMBH abundances from WD heating can be found in Fig.\ref{fig:catminlim}.

\subsection{Neutron Star Heating and Background X-Rays}
\label{sec:nsheating}

The effects of magnetic monopoles falling into NSs was explored thoroughly in  Ref.\cite{Kolb:1984yw}.  Magnetic monopoles cause nearby neutrons to decay and release their mass energy as lighter particles.  The power radiated by all the monopole-catalyzed decays heat the NS, increasing its x-ray radiation.  Constraints come from limiting the diffuse x-ray signal to be below the observed soft x-ray background.  This same analysis can be applied to EMBHs.

This picture is complicated by annihilation effects and uncertainty around what lies at the NS core.  EMBH self annihilation reduces the number able to catalyze nucleon decay, and, therefore, the resulting x-ray signal.  EMBHs cannot annihilate in NSs with a type II superconducting core, whereas a core that transitions to a type I superconductor or to an exotic state of matter may offer little resistance to annihilation. This uncertainty will be discussed below.

The NS must first capture passing EMBHs.  They lose energy due to friction with the NS's electron Fermi gas \cite{1998AcPPB..29...25K}.  We estimate the stopping power using Eq.\eqref{fermifric}, where $E_f \sim 100$ MeV is the electron Fermi energy, $n_e\sim10^{36}$cm$^{-3}$ is the electron density, and $v$ is the velocity of the infalling EMBH which we take to be the escape velocity $\sim 0.3 $ \cite{1984NuPhB.236..255H}.  Numerically we find that a NS with a 10 km radius can capture the entire relevant mass range of EMBHs.

Once inside, an EMBH radiates with a luminosity
\begin{equation}
    L_{ns} = 2.5\times10^{24} \frac{\sigma_{cat}}{\text{bn}} \left(\frac{v_{ns}}{0.3}\right)\left(\frac{\rho_{ns}}{5\times10^{14}\text{g cm}^{-3}}\right) \text{GeV s}^{-1},
\end{equation}
 where $v_n\sim0.3$ is the neutron velocity, $\rho_{ns}\sim 5\times10^{14}$g is the core density, and $\sigma_{cat}$ is the catalysis cross section.  For illustrative purposes, we  show this bound assuming $\sigma_{cat}=\sigma_{\text{QCD}}$.  When annihilation effects are not relevant, the EMBHs have a combined luminosity of  $L_{tot} = N_{acc}L_{ns}$, where $N_{acc}$ is the total number of EMBHs accumulated.  When annihilation is fast compared to the accumulation rate, the EMBHs that actually remain in the NS all carry the same sign charge.  Their number is set by a random walk with $N_{acc}$ steps.  In this case $L_{tot} = \sqrt{N_{acc}}L_{ns}$.
We take
$N_{acc} = t_{ns}\Gamma_{ns}$, where  $t_{ns}$ is the age of the NS, and the EMBH accumulation rate is
\begin{equation}
\label{nsaccume}
    \Gamma_{ns} \approx  \left(\frac{\rho_{dm}^{MW}f}{M}\right)\pi r_{ns}^2\frac{v_{esc}^2}{v}\sim5 f \left( \frac{m}{g}\right)^{-1} s^{-1} = 1.4\times10^{17}f\left(\frac{m}{g}\right)^{-1} \text{Gyr}^{-1},
\end{equation}
for an NS with $r_{ns}=10$ km and an escape velocity $v_{esc} = 0.3$.
The energy output from all of the EMBHs give the NS a surface temperature 
\begin{equation}
    T_{ns} = \left(\frac{L_{tot}}{4\pi r_{ns}^2\sigma_b}\right),
\end{equation}
where $\sigma_b$ is the Stefan-Boltzmann constant.\\

\subsubsection{Bounds from Overheating Neutron Stars}
The bounds on EMBH abundance come from limiting the x-ray radiation from catalysis heated NSs. We follow the analysis in  Ref.\cite{Kolb:1984yw}.  A NS heated to a temperature $T_{ns}$ radiates photons with energy $E$ like a black body with a differential luminosity of  \cite{Kolb:1984yw}
\begin{equation}
    \frac{dL}{dE} = 2\times10^{36}\left(\frac{r_{ns}}{10\text{km}}\right)^2\frac{E^3}{e^{E/T_{ns}}-1} \text{ ergs KeV}^{-4} \text{s}^{-1}.
\end{equation}
The differential flux that reaches Earth is
\begin{equation}
    \frac{dF}{dE} = \frac{dL}{dE}\frac{e^{-\tau(l,E)}}{4\pi l^2},
\end{equation}
where $l$ is the distance to the NS and $\tau(l,E)$ is the absorption length for photons with energy $E$ in the Galaxy and can be found in   Ref.\cite{Kolb:1984yw}. 

The catalysis decay luminosity of an individual NS depends on the number of EMBHs it has accumulated, which depends on its age.  To get the total x-ray flux from all of the NSs, one must integrate their luminosities as a function of time over the range of possible ages.  For simplicity, we assume NSs in the Galaxy are produced at a constant rate over $\sim 10^{10}$ years, and have a local number density of $n_{ns}\sim 10^{-4}$pc$^{-3}$ \cite{Kolb:1984yw}.  As we will explain in Section \ref{sec:annihilation effect}, EMBHs may be free to annihilate in up to $90\%$ of NSs.  Therefore when calculating the total radiation from nearby ones, we assume $10\%$ accumulate all EMBHs that pass through them, while the other $90\%$ only retain EMBHs that do not annihilate. 

Accounting for absorption in the ISM, the total differential flux from all NSs out to a distance of $d_{ns}\sim \text{few kpc}$, about as far as the x-ray signal is expected to travel without scattering \cite{Kolb:1984yw},  is \begin{equation}
    \frac{dF}{dE}_{tot} =\frac{n_{ns} B(E)}{t_0}\int_0^{t_0}\frac{dL}{dE} dt_{ns},
\end{equation}
where $t_{ns}$  is the age of the NS, $t_0$ is 10 Gyr, and $B(E)$ is the integrated fraction of photons with energy $E$ that reach Earth:
\begin{equation}
    B(E) = \int^{d_{nd}}_0 e^{-\tau(l,E)}dl
\end{equation}  The total flux that reaches Earth will be compared against sounding rocket soft x-ray observations from Ref.\cite{1983ApJ...269..107M}.  The counting rate per frequency band for a given flux is
\begin{equation}
\Gamma_{band} = \int\frac{dF}{dE_{tot}}A_{band}(E) dE
\end{equation}
where $A_{band}(E)$ is the detector response function for each of the four different x-ray detection bands and can be found in Ref.\cite{Kolb:1984yw}.

In large regions of the sky the counting rate measured by Ref.\cite{1983ApJ...269..107M} was $\frac{1}{3}$ the all-sky average  \cite{Kolb:1984yw}.  As in  Ref.\cite{Kolb:1984yw}, we set limits by requiring the count rate of soft x-rays produced from radiating NSs to be less than this.   We use the weakest of the limits derived from the four bands.  The resulting bounds are shown in Fig.\ref{fig:catQCD}.

\subsubsection{Annihilation Effects and the Superconducting Core}
\label{sec:annihilation effect}

Whether EMBHs can annihilate in a NS depends on the properties of its inner core.   The inner $\sim$5 km is thought to be occupied by a type II superconductor   \cite{Haskell:2017lkl}.   When an EMBH falls into a NS, it sinks  into the superconducting surface, which produces 2 flux tubes around the EMBH for each unit of dirac magnetic charge, $q_d = \frac{1}{2e}$, it carries.  Each flux tube carries a magnetic flux of $\frac{\pi}{e}$ and an energy per unit length of $\sim 2\times10^{11}$ GeV cm$^{-1}$ \cite{PhysRevD.33.2084}.  This acts as a tension pulling the EMBH back toward the surface of the superconducting core.

The EMBHs form a suspended surface where gravitational and tension forces balance, $\sim0.6$ km above the center of the NS independent of the EMBH mass.  In agreement with  Ref.\cite{bai2020phenomenology}, we find that annihilation between suspended EMBHs is too slow to alter the radiation bounds.   However, NSs are poorly understood at these depths and may transition to a new state which does not support the EMBH surface and prevent annihilation (such as a type I superconductor  \cite{Sedrakian_1997, Haskell_2018} or some exotic state of matter  \cite{Landry_2020}).    Any such transition would reduce the number of EMBHs available to catalyze neutron decay and severely weaken the bounds. 

NS cores require densities above  $\sim7\times10^{14}$g cm$^{-3}$ to transition  \cite{Sedrakian_1997}.  While the equation of state and mass range that avoid exotic transitions remains an area of active research  \cite{Landry_2020}, lighter NSs are expected to have less dense cores that do not transition \cite{emanuele}.  Based on modeling in  Ref.\cite{Silva_2016}, we estimate that NSs with masses $\lesssim 1.1 M_{\odot}$ can reasonably be treated as having a core density below the transition threshold.   Based on the available mass measurements for NSs  \cite{doi:10.1146/annurev-astro-081915-023322}, we estimate that $\sim10\%$ have masses below $\sim 1.1M_{\odot}$.  These NSs do not permit EMBH annihilation and so keep those they accumulate.  The remaining $90\%$ amass EMBHs at a random walk rate proportional to the square root of the number that pass through them.  We make the conservative assumption that annihilation is fast whenever it is possible, as this gives the lowest estimate for the number of EMBHs accumulated, the smallest NS x-ray luminosity, and the weakest bounds.  

At high masses ($M \gtrsim 10^9$ g), a single EMBH can cause a NS to overheat, so annihilation suppression no longer matters.  Only few EMBHs with $10^7$g $<M<10^9$g are needed to overheat a NSs, so $N_{acc}$ is not much smaller than $\sqrt{N_{acc}}$, and all NSs contribute to the bound.  The bound on EMBHs with $M<10^7$g is dominated by radiation from non-annihilating NSs.

\subsection{Proton Decay in the Sun, Neutrinos and SuperK}
\label{sec:inSun}

Proton decay in the Sun has been constrained by neutrino observations at Super Kamiokande  \cite{Ueno:2012zua}.  Most of the EMBHs we focus on would become trapped while passing through the Sun and cause proton decay once inside.  Heavier EMBHs get caught in the convection zone, while those lighter than about 300 g make it to the core and eventually annihilate.  Bounds  come from the steady-state number in the Sun.  These bounds are weaker than those derived for NSs and WDs when the catalysis cross section is held constant.  However, some enhancements to the cross section from EMBH-proton bound states \cite{Bracci:1983fe,Olaussen:1983bm} or angular momentum effects \cite{PhysRevLett.50.1901} can make the solar neutrino bound stronger than the WD heating bound for some masses.  Details of this constraint are included for reference when considering specific EMBH models with specific $\sigma_{cat}$.  As a toy example, we will assume $\sigma_{cat} =\sigma_{QCD}$ for the remainder of this section.   We also note that   Ref.\cite{bai2020phenomenology} derives stronger solar bounds by considering neutrino emission from EMBHs that annihilate in the solar core.  These  bounds should apply to EMBHs which are light enough; however, Ref.\cite{bai2020phenomenology} does not consider how those heavier than 300 g get trapped in the convection zone which suppresses their annihilation rate and weakens these constraints.

 Super Kamiokande is a water Cherenkov detector sensitive to  neutrinos with energies above 5.5 MeV  \cite{Abe_2011}.  About half of proton decays produce charged pions, which themselves dominantly decay via $\pi^+\rightarrow\mu^+ +\nu_{\mu}\rightarrow e^+ +\nu_e +\bar{\nu}_{\mu} + \nu_{\mu}  $, producing three neutrinos with O(10 MeV) energies  \cite{Ueno:2012zua}.     The analysis in  Ref.\cite{Ueno:2012zua} limits the flux of  neutrinos resulting from proton catalysis in the Sun with 90$\%$ certainty to $I_{90}<166.6$ cm$^{-2}$ s$^{-1}$.  
This limits the proton decay rate in the Sun to be less than
\begin{equation}
\label{eq:solarbound}
    F_p^{max} = \frac{4\pi d^2 I_{90}}{b_{\pi^+}(1-a_{\pi^+})} = 4\times10^{29} s^{-1},
\end{equation}
where $b_{\pi^+} \sim 0.5$ is the branching fraction of protons into $\pi^+$,  $a_{\pi^+}\sim 0.2$ is the probability that a pion gets absorbed by the solar core  \cite{Ueno:2012zua}, and $d$ is the Earth-Sun distance.

An EMBH passing through the Sun will lose energy as described in Eq.\eqref{eqn:friction}.  Since the average velocity is $v\simeq 10^{-3}$ and the escape velocity from the solar surface is $v_{esc} \simeq 2\times10^{-3}$, the EMBH needs to lose much of its kinetic energy to become gravitationally bound to the Sun.  Taking an average internal temperature for the Sun of $T= 2\times 10^6 K$ in Eq.\eqref{eqn:friction}, one finds this requires $M\gtrsim 0.5$ g.  The Sun accumulates such EMBHs at the (slightly) Sommerfield enhanced rate of  
\begin{equation}
    \Gamma_{\odot} \sim \pi r_{\odot}^2\left(\frac{\rho_{dm}^{MW}f}{M}\right)\left(1+\frac{v_{esc}^2}{v^2}\right)v = 4\times10^{22} \text{Gyr}^{-1}f\left(\frac{M}{ \text{ g}}\right)^{-1},
\end{equation}
where $r_{\odot}\simeq 7\times 10^5$ km is the solar radius.

Heavier EMBHs will get trapped in the convection zone, which makes up the outer third of the Sun.  The temperature varies from $2\times10^6$ K at the base of the convective zone to 5700 K at the surface.  The convective motions have velocities around $10^{-6}$  \cite{1964ApJ...140.1120S}.  Using the solar mass profile $M_\odot(x)$  \cite{1971ApJ...170..157A}, we add the term $G M M_\odot(x)/(r_\odot -x)^2$ to Eq.\eqref{eqn:friction} to account for the change in gravitational potential and numerically integrate the energy loss as an EMBH traverses the Sun.  The velocity drops below the convection value if $M \gtrsim 300$ g. Thus, this range of masses will be trapped in the convection region.

One can check that annihilation within the convection region is unimportant.  EMBHs trapped in the convection region will move with stellar convection currents and annihilate at a rate $ \Gamma \sim N_{\odot} n_{\odot} \sigma v_{th}$, where $N_{\odot}$ and $n_{\odot}$ are the number and number density of EMBHs accumulated in the convection region, $v_{th}$ is the thermal velocity, and  $\sigma$ is a cross section defined by the region where two EMBHs approaching each other at free-fall velocity can meet within the time a convection flow would cross the convection region $\sim 7\times10^5$s (a rough `coherence' time for the flow).    $\Gamma$ is never large enough to significantly alter the number of EMBHs accumulated in the Sun, so we neglect its contribution to these constraints.

The proton catalysis rate is
\begin{equation}
\label{eqn:catSun}
    F_p = \Gamma_{\odot}t_{\odot}n_p v_p \sigma_{\text{cat}},
\end{equation}
where $t_{\odot}=4.6$ Gyr is the age of the Sun, $n_p$ is the proton number density set by the local density, $0.2$ g cm$^{-3}$, and $v_p\sim10^{-4}$ is the local proton thermal velocity.  For illustrative purposes, we set $\sigma_{\text{cat}}=\sigma_{\text{QCD}}$.  Comparing this rate with the current bound \eqref{eq:solarbound} gives our constraint for $M>300$ g.

EMBHs with $M<300$ g pass through the convection region and sink to the core.  The vast majority that reach the core will ultimately annihilate.  Only those that have not yet done so can contribute to proton catalysis.  The total number in the Sun at any time, $N_{tot}$, is made up of those that have not yet annihilated because they are either in the process of sinking to the core ($N_{sink}$) or part of a charge imbalance due to the random walk of charge accumulation ($N_{ran}$).  There could additionally be a delay in annihilation if there is a strong enough magnetic field in the core, but we will ignore this possibility to set a conservative bound.  Thus, $N_{tot}=N_{sink}+N_{rand}$.  

We estimate the number of EMBHs in transit as:
\begin{equation}
    N_{sink} = \Gamma_{\odot}t_{sink}\, ,
\end{equation} 
where the sinking time $t_{sink}$ can be computed numerically.  Using estimated temperature and density profiles in the Sun  \cite{1971ApJ...170..157A}, we find
\begin{equation}
    t_{sink}\sim 10^3 s \left(\frac{M}{g}\right)
\end{equation}
for $M>0.5$ g, implying $N_{sink} \sim 10^9 f$.  On the other hand, the number accumulated from random asymmetries in the signs of the EMBHs passing through the Sun over its lifetime $t_\odot$ is
\begin{equation}
    N_{rand} = \sqrt{\Gamma_{\odot}t_{\odot}} \simeq 4 \times 10^{11} \sqrt{f} \left( 
    \frac{M}{1\, {\rm g}}\right)^{-\frac{1}{2}}.
\end{equation}
The EMBHs actively sinking toward the core and those in the innermost region of the star are in different environments and have different catalysis rates.
We estimate the outer core where EMBHs are sinking to have $T_{out}\sim7\times10^6$ K and $\rho_{out}\sim50$ g cm$^{-3}$, and the inner core where the random walk EMBHs reside to have $T_{in}\sim 16\times10^6$ K and $\rho_{in}\sim 150$ g cm$^{-3}$.  Using this information and Eq.\eqref{eqn:catSun}, one can find the catalysis rates for both sinking and random walk EMBHs.  Combining these gives the total proton decay rate, which must be less than $F_p^{max}$ given by Super K. The resulting bounds are shown in Fig.\ref{fig:catQCD}.

One might worry that the accumulated random walk magnetic charge could be large enough to alter the dynamics in the stellar core.  The maximum possible random walk charge for EMBHs that accumulate here would produce a magnetic field $\sim 10^{-6}$G in the lower convection region.  This is negligible compared to the magnetic field already present, which could be as strong as 100 G \cite{2007A&A...461..295A}, and is thus unlikely to influence stellar dynamics.

\subsection{Direct Constraints}
\label{sec:MACROcat}

The magnetic monopole detector MACRO, described in Section \ref{sec:MACROnocat}, is sensitive to nucleon decay effects.  Nuclear decay catalyzed by a traversing monopole produces extra radiation which changes the detector's sensitivity.  MACRO produced separate analyses and flux bounds for monopoles that do \cite{Ambrosio:2002qu} and do not catalyze nucleon decay  \cite{Ambrosio:2002qq}.  In the catalysis case, the flux limit from MACRO data depends on the catalysis cross section.  The flux limit for a magnetic charge moving at $v\sim10^{-3}$, $F(\sigma_{cat})$, is presented for magnetic charges with catalysis cross sections up to $\sigma_{cat}=10^{-24}$cm$^{2}$ in Ref.\cite{Ambrosio:2002qu}. Increasing the catalysis cross section weakens MACRO's ability to verify EMBH events  \cite{Ambrosio:2002qu}.  We take this cross section and the corresponding flux limit $F\lesssim10^{-15}$ cm$^{-2}$ s$^{-1}$ sr$^{-1}$, as it is the  most conservative limit considered in  Ref.\cite{Ambrosio:2002qu}.  This gives the below constraint, which is presented in Fig.\ref{fig:catQCD}.

\begin{equation}
    775<\left(\frac{m}{g}\right)^{-1}f.
\end{equation}

\section{Sub-Extremal Magnetic Black Holes}
\label{sec:nonextremal}
This work so far has focused on EMBHs in a mass range, ($M<10^{15}$g), in which MBHs must be extremal to stable over the lifetime of the Universe. If the $Q$ enhancement to the radiation rate of leptons and fermions described in  Ref.\cite{Maldacena:2020skw}  is present, then sub-extremal MBHs only survive to the present if their surface temperature is too low to effectively radiate electrons and positrons ($M \gtrsim 10^{17}$g)  \cite{Maldacena:2020skw}.  The sections below explore the behavior of and abundance constraints on MBHs heavy enough to be stable over the lifetime of the Universe even when sub-extremal ($M>10^{17}$g). Some of the previously discussed constraints on EMBHs continue to apply at these high masses and  also apply to sub-extremal MBHs.   Heavy MBHs are constrained by their interactions with warm ionized gas clouds in the Milky Way and with WDs.  Some can cause NSs to collapse into black holes. For convenience, we describe the charge of an MBH using $q$, a fractional value of the charge compared to the extremal charge, $Q_{ext}$, of a black hole with mass $M$.
\begin{equation}
    q\equiv\frac{Q}{Q_{ext}}=\frac{Q m_{P}}{M}.
\end{equation}
\\
\subsection{Gas Heating}
\label{sec: Gas Heating}

Generalizing the constraints on EMBH abundance due to overheating warm gas clouds (see Section \ref{warmgas})  for MBHs gives:
\begin{equation}
        q^2f<3 \left(\frac{M}{10^9g}\right)^{-1}.
\end{equation}
This bound cuts off when MBHs either pass through WIM clouds too rarely to alter their behavior or lose a significant fraction of their kinetic energy to friction with the interstellar medium.   Based on Eq.\eqref{eqn:friction}, an MBH with  $M\gtrsim3\times10^{21}$ g may deposit the total energy needed to overheat the entire cloud in a single passage.  The energy is initially deposited in a small region defined by the plasma attenuation length $l\sim 16$ km.  Depositing the energy needed to heat a $\sim2$ pc cloud into a cylindrical cross section with a radius of $\sim$16 km will heat the region to extreme temperatures, causing it to expand supersonically.  This would significantly alter the cloud's structure and behavior within its sound crossing time, $\sim 5\times 10^5$ yr for typical warm clouds \cite{1977ApJ...218..148M}.  Natural processes such as cooling and supernova shocks evolve the WIM over timescales of $10^6$ yr \cite{1977ApJ...218..148M}.  MBHs must not disrupt WIM clouds more quickly than naturally observed processes.  We limit the abundance of MBHs with $M>3\times10^{21}$ g, so that typical WIM clouds encounter them less than once per $10^5$ yr. 

This bound also cuts off when MBHs lose an $\mathcal{O}(1)$ fraction of their kinetic energy to friction with the WIM within $\sim 10$ Gyr.  Losing this much kinetic energy causes them to sink toward the Galactic center. There may be interesting phenomenology around this, but we have not found new constraints that arise from it.   Modifying Eq.\eqref{eqn:Elossgas} gives
\begin{equation}
    \frac{dE}{dt} = 9\times10^{-3}  q^2 \left(\frac{M}{10^9\text{ g}}\right)^2 \text{erg s}^{-1}
\end{equation}
for MBHs passing through warm ionized regions.  They lose energy most efficiently in the WIM, which fills $1\%$ of the Galactic disk \cite{2011piim.book.....D}.  Most dark matter in the Galaxy, $\sim 90\%$, sits outside of the Galactic disk in the dark matter halo.  We assume then that MBHs moving along virial orbits around the Galactic center spend $\sim 10\%$ of their time in the Galactic disk.  Using all of this, we estimate that MBHs lose an $\mathcal{O}(1)$ fraction of their kinetic energy within 10 Gyr if 
\begin{equation}
    \left(\frac{M}{10^{19}\text{ g}}\right) q^2\gtrsim 1.6.
\end{equation}
The gas heating bound cuts off where this is saturated.  The bounds for MBHs with different parameters are shown in Fig.\ref{heavybh}.
\\
\subsection{White Dwarf Consumption}

MBHs with $M>10^{17}$g can get trapped inside of a 1.2M$_{\odot}$ WD and consume it within 1 Gyr.  Their abundance is constrained by the existence of hundreds of WDs older than a Gyr and with masses $\sim1.2M_{\odot}$ \cite{2017ASPC..509....3D}.  These WDs yield the strongest bounds because they are both abundant and dense enough to capture a wide range of MBHs.   Uncharged black holes simply pass through a WD, while MBHs are trapped by friction with the Fermi gas.  MBHs experience the same frictional forces in a WD as EMBHs (described in Eqs.\eqref{fermifric} and \eqref{eqn:friction}) but reduced by $q^2$.   Using this and the 1.2 M$_{\odot}$ WD density profiles derived in  \cite{cococubed}, we find that all black holes that satisfy  
\begin{equation}
\label{qmin}
    q\gtrsim3\times10^{-16}\left(\frac{M}{10^{17}\text{g}}\right)^{-1/2}
\end{equation}
will get trapped in any 1.2M$_{\odot}$ WD they pass through.

Estimates in  Ref.\cite{Fedderke_2020} indicate that an uncharged black hole should be able to consume an entire WD within
\begin{equation}
    t\sim\frac{c_s^3m_{P}^4}{4\pi \lambda\rho_{c}}\left(\frac{1}{M}-\frac{1}{m_{WD}}\right)\sim 32 \text{ Myr } \frac{10^{17}\text{g}}{M},
\end{equation}
where $c_s\sim 3\times10^{-2}$ is the speed of sound in a 1.2 M$_{\odot}$ WD, $\lambda$ is an $\mathcal{O}$(1) constant presented in Ref.\cite{Fedderke_2020}, $\rho_{c}\sim10^8$g cm$^{-3}$ is the density at the core of a 1.2 M$_{\odot}$ WD, and $m_{WD}$ is the mass of said WD  \cite{Fedderke_2020}.  One might worry that interactions with the large magnetic field around MBHs slows accretion or that MBHs cannot absorb electrically charged particles without becoming super-extremal.  While the magnetic field around an MBH can repel charged particles before they reach the event horizon  \cite{Grunau_2011}, the outward flow of repelled material would be resisted by pressure from the WD.  The pressure from $\rho_{c}\sim10^8$g cm$^{-3}$ of material moving away from the MBH at $c_s$ is approximately four times greater than the ambient pressure in the WD core. Balancing forces, $\rho_{wd}$ is reduced by at most a factor of $\sim$4 in the area around the MBH, increasing the accretion time by up to a factor of $\sim$4.  All MBHs with $M>10^{17}$g will still be able to consume an entire 1.2 M$_{\odot}$ WD within a Gyr.  EMBHs in this mass range do not become super-extremal upon absorbing a single electrically charged particle because the magnetic and electric charges add in quadrature.

\begin{figure}
    \centering
    \includegraphics[scale=0.43]{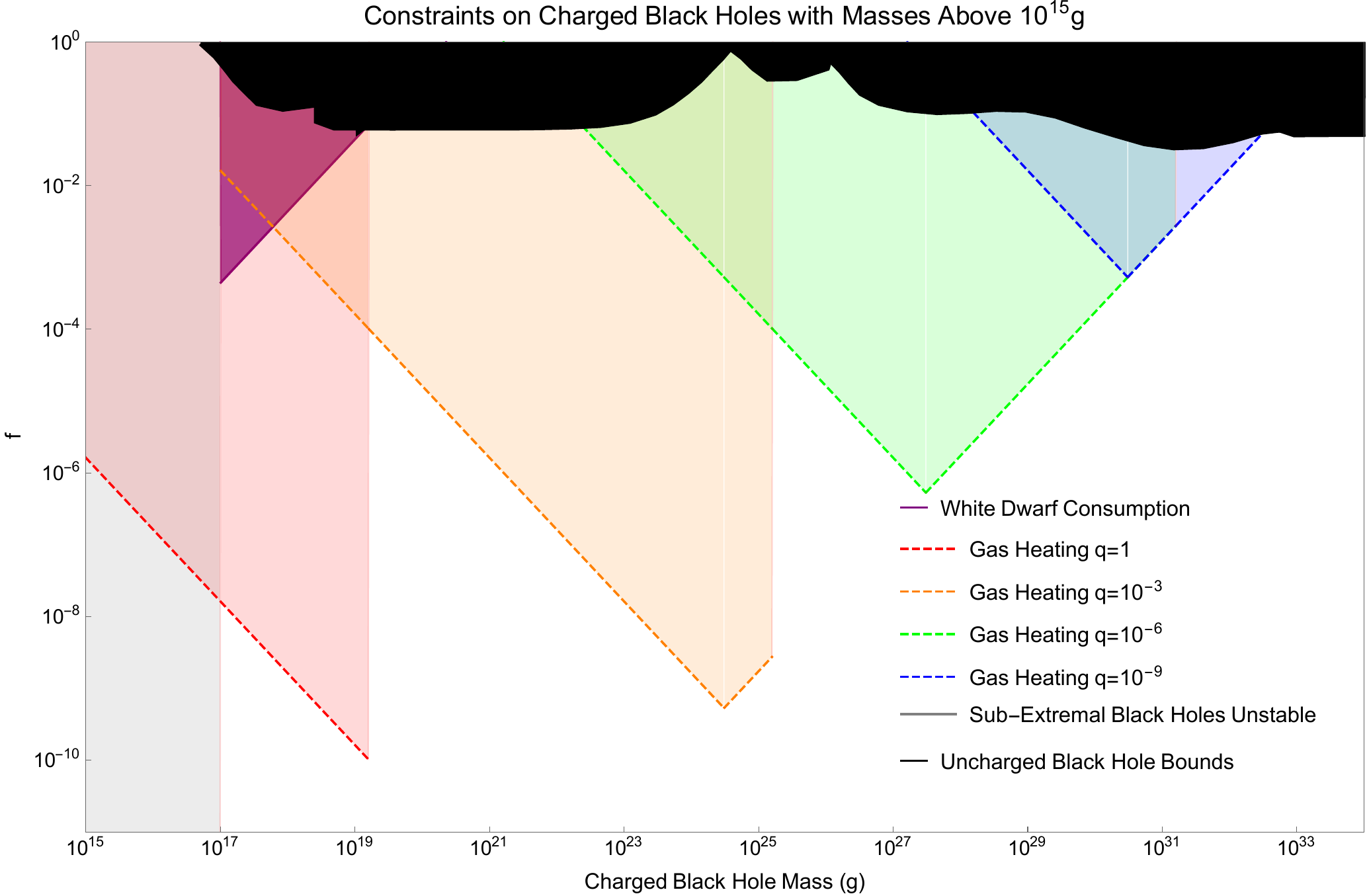}
    \caption{These are the constraints on MBHs with $M>10^{15}$g.  The purple region is excluded for all MBHs that satisfy Eq.\eqref{qmin} due to WD consumption. The dashed lines represent bounds from overheating WIM clouds in the Milky Way.  Each color traces the bound for MBHs with a different value of $q$.  This bound tilts upward when a typical WIM cloud only encounters an MBH once every $10^5$ years.  The bound cuts off when MBHs lose an $\mathcal{O}$(1) fraction of their kinetic energy to friction with the WIM clouds within 10 Gyr.  MBHs are unstable on cosmological timescales if $q\neq1$ in the gray region \cite{Maldacena:2020skw}. Bounds on uncharged black holes applicable to MBHs are shown in black \cite{carr2020constraints}.}.  
    \label{heavybh}
\end{figure}

Typical 1.2 M$_{\odot}$ WDs must not accumulate 1 MBH capable of consuming it within 1 Gyr.  This gives
\begin{equation}
    f<10^{-4}\left(\frac{M}{10^{17}\text{g}}\right)
\end{equation}
for all MBHs that satisfy Eq.\eqref{qmin}.  This constraint is shown in Fig.\ref{heavybh}

\subsection{Neutron Star Collapse}  The recent LIGO/Virgo observation of a 2.6 M$_{\odot}$ compact object occupying the mass-gap between the theoretical maximum mass for NSs and minimum mass for stellar black holes \cite{Abbott:2020khf} has raised questions about how such an object could form.  Ref.\cite{dasgupta2020low} suggests that a low mass black hole, perhaps the object observed by LIGO/Virgo, could result from a NS absorbing non-annihilating dark matter and collapsing into a black hole.  Some MBHs can get trapped in NSs due to friction with the Fermi gas and cause this collapse.

The minimum accretion rate for a black hole in a degenerate neutron gas with a stiff equation of state (the conditions of a NS core) was found \cite{richards2021relativistic} to be:
\begin{equation}
    \dot{M} = 3\times 10^4 \left(\frac{M_{\odot}}{\text{s}}\right)\frac{M}{M_{\odot}}
\end{equation}
An MBH can therefore consume a $\sim 2 M_{\odot}$ NS within 
\begin{equation}
    t =0.8 \text{ Gyr} \left(\frac{M}{10^{22}\text{g}}\right)^{-2}.
\end{equation}
Those with masses above $\sim 10^{22}$ g can easily consume a NS within 1 Gyr, causing it to collapse into a mass-gap black hole.

Only MBHs that get caught in NSs can consume them.  Using Eq.\eqref{WDfric}, we find NSs with $E_f \sim 100$ MeV,  $n_e\sim10^{36}$cm$^{-3}$, $v_{esc}\sim0.3$  and $r=10$ km   \cite{1984NuPhB.236..255H} will trap any MBH that passes through which satisfies 
\begin{equation}
\label{eatNS}
    q\geq  10^{-16}\left(\frac{M}{10^{22}\text{g}}\right)^{-1/2}.
\end{equation} Using Eq.\eqref{nsaccume} one finds the probability that a NS captures one of these MBHs is $1.4\times 10^{-5} f \left(10^{22}\text{g}/M\right)\left(\frac{T}{\text{Gyr}}\right)$, where $T$ is the age of the NS.  Clearly, MBHs heavy enough to collapse a NS $(M>10^{22}\text{ g})$, and that satisfy Eq.\eqref{eatNS}, could never be abundant enough to observably alter the NS abundance.    Regardless, they offer a new way to produce black holes in the mass gap range.

\section{Discussion}
\label{sec:Conclusions}
\subsection{Generalized Mass Functions}
Throughout this paper we have assumed that EMBHs have a monochromatic mass function.  It is worth exploring how the bounds described in the above sections transform for EMBHs with non-monochromatic mass functions, as these commonly arise from models of primordial black hole formation.  We outline how one would apply the constraints described in this paper to EMBHs with a general mass function below.  We describe the mass function using $\frac{dn}{dM}(M)$, which gives the differential number density of EMBHs as a function of mass.

The constraint from WIM cloud heating in Section \ref{warmgas} is estimated by comparing the total heat generated by many EMBHs passing through WIM clouds in the Milky Way to the observed heating and cooling rates of said clouds.  The heating rate for EMBHs with a general mass function can be found by multiplying the heating rate per EMBH,  Eq.\eqref{eqn:Elossgas}, and $\frac{dn}{dM}$, and then integrating over all relevant masses. This can be compared against the observed cooling rate to obtain bounds.

The constraints from EMBH annihilation in  WDs leading to runaway fusion and supernovae is  described in Section \ref{sec:wdann}.  The total mass of EMBHs accumulated in a WD depends on the local density of dark matter and not the mass of the EMBHs themselves.  Additionally, a set accumulated mass of EMBHs, $M_{Th}$, is needed to overcome the core magnetic field and start annihilation (see Table \ref{tab:table1}).  The annihilation bound is sensitive to the whether EMBHs have enough mass to provide the energy needed to initiate runaway fusion, and whether they are light enough to sink to the core within 1 Gyr.  The bounds from  WD J0551+4135 can be straightforwardly adapted for general EMBH mass functions because all EMBHs lighter than $4\times 10^{12}$g can sink to the core and trigger runaway fusion within 1 Gyr.   WD J0551+4135 must not accumulate a group of at least six EMBHs, each lighter than $4\times 10^{12}$g, whose combined mass exceeds $M_{Th}$ within 1 Gyr.   Not all EMBHs supply enough energy when they merge to trigger runaway fusion in 1 $M_{\odot}$ WDs.  This complicates estimating the bounds because the non-extremal black hole produced in an EMBH merger will only radiate off the mass of the smaller EMBH. One can simplify this estimate by considering two different sets of EMBHs.  The first is those with $4\times 10^{12}>M\gtrsim M_{Th}$.  These will trigger runaway fusion once the WD accumulates 6 EMBHs whose combined mass exceeds $M_{Th}$.   The second set is EMBHs with $M\ll M_{Th}$.  If we make the simplifying assumptions that all EMBHs in this group  pair and annihilate exactly once, that annihilations are fast compared to the time needed to accumulate $M_{Th}$ of EMBHs, and that the distribution of EMBHs in the WD matches the true distribution, then one could estimate the number of times the WD would need to accumulate $M_{Th}$ of EMBHs before it would be expected that one of the annihilating pairs contains two EMBHs heavy enough to trigger runaway fusion.  The faster of the two processes can be used to set the bound. 

Generalizing the constraints on EMBH abundance due to gamma-rays produced when EMBH binaries annihilate (Section \ref{sec:binarymerger}) requires modifying the derivation for the EMBH merger rate presented in Appendix \ref{sec: app merge}.  The details of the original and modified derivation are described there.  Broadly, the merger rate comes considering a collection of cold EMBHs whose separations in the early Universe are described by a Poisson distribution.  These initial separations dictate which EMBHs form binaries, the semi-major axes and eccentricities of these binaries, and the time they take to merge.  The merger rate at any time $t$ comes from integrating over all binaries whose semi-major axis and eccentricity lead them to merge after $t$ has passed.  When the mass spectrum is not monochromatic, one must include terms in the distribution of binary parameters that account for the distribution of EMBH masses.  The merger rate then comes from integrating this over the space of EMBH masses, binary semi-major axes, and eccentricities which produce binaries that merge after $t$ has passed. Each merger remnant will radiate off the mass of the lighter EMBH in the binary before becoming extremal again.  This amount of radiation per merger, along with the modified merger rate described in Appendix \ref{sec: app merge} can be used to estimate the total diffuse gamma-ray signal produced by binary mergers, which can be compared to the observed diffuse extra galactic gamma-ray signal observed by FERMI-LAT. 

The existence of large magnetic fields in both the Milky Way and in galaxy clusters was used to limit EMBH abundance in Section \ref{sec:destruction of Mfield}.  Too many EMBHs would damage the Galactic magnetic field by removing energy from it and would screen the large coherent magnetic fields observed in galaxy clusters.  Both effects are mass independent, and limit the total fraction of dark matter that can be made of EMBHs of any mass.    

The MACRO monopole detector  set upper limits on the flux of magnetic charges near the Earth (see Section \ref{sec:MACROnocat}).  One can use this to constrain EMBHs with arbitrary mass distributions by comparing the expected flux of EMBHs, given by $\frac{1}{4\pi}v\int\frac{dn}{dM}dM$, where $v\sim 10^{-3}$ is the local EMBH velocity, to the MACRO flux.

The way EMBH catalysis of nucleon decay affects WD annihilation bounds is explored in Section \ref{sec:convect}.  This is somewhat more complicated for EMBHs with non-monochromatic distributions, as the catalysis cross section typically varies with EMBH mass.  The heat produced by catalysis can cause convection in WD cores which prevents EMBH annihilation.  If the annihilation bounds break down, the EMBH abundance is constrained by comparing the predicted luminosity from EMBHs catalysing nucleon decay in WDs to that observed in old dim WDs. One can determine whether the annihilation bounds stand by comparing $L_{th}$, the luminosity from catalysis decays needed to start convection in WDs, to $L_{M}$, the maximum plausible luminosity from a collection of EMBHs in the WD.  When the typical EMBH masses are small compared to $M_{Th}$, $L_{M}$ can be determined by integrating $\int\frac{M_{Th}}{n}\frac{dn}{dM}\sigma_M\rho_{c}v_{th}dM$.  Here $n$ is the number density of EMBHs, $\sigma_M$ is the maximum plausible catalysis cross section, defined in Eq.\eqref{eq:Sigmax}, $\rho_{c}$ is the density at the WD core, and $v_{th}$ is the thermal velocity of nuclei there.  As in the monochromatic case, the annihilation bounds remain in tact unless $L_{th}<L_{M}$ at all WD cooling temperatures.  When annihilation bounds break down, one can estimate the average cross section per unit of EMBH mass using $\frac{L_{th}}{v_{th}M_{Th}\rho_c}$.  This can be used to estimate the luminosity EMBHs would produce in an old dim WD and compared to the observed luminosities to get constraints.  If the typical EMBH mass is large compared to $M_{Th}$, then only 6 are needed to cause the WD to explode.  In this case, one can estimate $L_M$ as the expectation value for the luminosity of 6 randomly selected EMBHs radiating at $\sigma_M$.  If the annihilation bound breaks down, one can find the heating bound by using the same general method described for light EMBHs, though using $L_{th}/{6 v_{th}\rho_c}$ as the average catalysis cross section.

Section \ref{sec:nsheating} describes the EMBH abundance constraints from nucleon catalysis in many nearby NSs producing too much x-ray radiation.  The catalysis cross section we used to demonstrate this constraint, $\sigma_{QCD}$, is proportional to the EMBH mass.  With this cross section, the nucleon decay rate and the resulting bound relies on the total mass of EMBHs accumulated in NSs and not generally on the mass of the specific EMBHs involved. The bound becomes mass dependent once EMBHs are heavy enough ($M>10^7$ g) that one or a few of them cause NSs to produce too much radiation.  At higher masses ($M>10^9$ g) individual EMBHs captured in a fraction of all NSs can saturate the limits on  x-ray radiation. One can assume the EMBH mass distribution across all NS reflects the true distribution.  In this case, one can estimate the fraction of NSs which would acquire a given luminosity based on the probability that they accumulate a certain number of EMBHs of a given mass.  Integrating this over all masses would give the expected combined luminosity from all NSs, which could then be compared to observed x-ray measurements to get constraints.  

Section \ref{sec:inSun} describes the constraints that emerge from EMBH induced proton decay in the Sun producing too many energetic neutrinos.  As in Section \ref{sec:nsheating}, we demonstrate this bound using $\sigma_{QCD}$ as the catalysis cross section, which makes it dependent on the total mass of EMBHs trapped in the Sun, and not the mass of said EMBHs.  Those lighter than 300 g annihilate too quickly to produce observable decays, while those heavier than 300 g get trapped in the convection zone where annihilation is strongly suppressed.    Generalized bounds come from limiting the fraction of EMBHs heavier than 300 g so that the proton decay rate stays below observed limits.

\subsection{Seeding Magnetic Fields}
One possibly interesting area we did not address is the connection between EMBHs and primordial magnetic fields.  It has been suggested that primordial magnetic fields present just before recombination could reduce the Hubble tension by altering the way charged particles cluster  \cite{jedamzik2020relieving}. Additionally, many models of galactic magnetic field formation rely on the presence of a small seed field around the time of galaxy collapse \cite{Beck:2013bxa}.  The seed fields needed to generate the $\mu G$ magnetic fields observed in galaxies today vary with the particular model, but reach as low as $10^{-30}$G fields coherent over distances of $\gtrsim 100$ pc \cite{PhysRevD.60.021301}.  EMBHs and MBHs carry high magnetic charges and can generate reasonably strong magnetic fields coherent over the average separation between them.  Taking the typical EMBH separation, one can estimate the scale of the typical magnetic fields they produce:
\begin{equation}
    B\sim\frac{4}{3}\left(\frac{4\pi\rho_{dm}f\Delta}{3M}\right)^{2/3}Q(1+z)^2\sim1.5*10^{-23} \text{ G }\times (f\Delta)^{2/3}q\left(\frac{M}{g}\right)^{1/3}(1+z),
\end{equation}
where $\Delta$ is the local EMBH overdensity.  Such a field would be coherent over length scales
\begin{equation}
    l\sim\left(\frac{3M}{4\pi\rho_{dm}f\Delta}\right)^{1/3}\sim 1.5*10^{-9}\text{pc }\times \left(\frac{M}{g}\right)^{1/3}\frac{(f\Delta)^{-1/3}}{(1+z)}.
\end{equation}
There are some unconstrained regions of parameter space where MBHs could play a role in galactic magnetic field formation.  This idea is explored further in Ref.\cite{Araya_2021}.

\section{Conclusion}
\label{sec:Conclusion}

In this work, we have extensively explored the phenomenology of EMBHs and stable MBHs.  We have placed stringent bounds on their abundances based on their ability to destroy WDs either by initiating supernovae  (EMBHs with M < $4\times10^{12}$ g) or by consuming them (MBHs with M> $10^{17}$ g). We also derived bounds which apply to both extremal and non-extremal MBHs based on overheating WIM clouds in the Milky Way.  We described how EMBHs merge and the observable emission that results from these mergers.  We provided minimal model-independent constraints for EMBHs that catalyze nucleon decay, while leaving a precise calculation of the catalysis cross section for future work.  As in Refs.\cite{bai2020phenomenology,ghosh2020astrophysical}, we do not consider a detailed exploration of how EMBHs form.

There is still much to understand about EMBHs.  How exactly low mass EMBHs accrete, whether nucleon catalysis happens, and the exact details of such a process are not yet understood.  While it is quite clear that EMBHs cannot constitute a meaningful fraction of the dark matter, they remain long-lived, well-motivated, very simple extensions of the Standard Model.  They offer new mechanisms for heating Galactic plasma, producing gamma-rays, initiating supernovae, and generating NS mass black holes. Their behavior also gives us insight into more exotic objects, such as higher dimensional black holes, which carry magnetic charge \cite{Bah:2020ogh}. EMBHs remain interesting tools to consider for many current and future astrophysical questions.

\vspace{.2in}
\textbf{Acknowledgements}
We would like to thank Emanuele Berti, Rob Farmer, Michael Fedderke, Ely Kovetz, Michael Shara, and Raman Sundrum for their insightful conversations and guidance. Special thanks to Leandro Althaus for sharing their detailed analysis of WD interior compositions. This work was supported in part by the National Science Foundation under grant PHY-1818899.

\appendix
\section{Magnetic Friction with a Fermi Gas at Finite Temperatures}
\label{sec:appfermi}
The friction between a magnetic charge and an electron Fermi gas at finite temperature can be found by modifying the derivation presented in Ref.\cite{PhysRevD.26.2347}.  The energy of electrons that scatter off a passing magnetic charge will change by $\Delta E= pv(\text{cos}\theta-\text{cos}\theta')$,  where $\theta$ is the angle between the velocity of the incoming electron and the magnetic charge velocity $v$, $\theta'$ is the angle between the scattered electron's velocity and $v$, and $p$ is the momentum of the electron.  The energy given to a magnetic charge from all of the scattering electrons can be found by integrating:
\begin{equation}
\label{startingfric}
    \frac{dE}{dx} = \int_{E,\theta,\theta',\Psi,\phi} n(E)\left(1-f(E+\Delta E)\right)\frac{d\sigma}{d\Omega}\Delta E\frac{\text{sin}\theta\text{sin}\theta'\text{sin}\Psi d\theta d\theta'd\Psi d\phi}{4\pi}.
\end{equation}
Here
\begin{equation}
    n(E) = \frac{m_e}{\pi^2}\sqrt{2m_e E}\times\frac{1}{\text{exp}\left(\frac{E-E_f}{T}\right)+1}
\end{equation}
is the number density of electrons with energy $E$ in a Fermi gas at temperature $T$ and with a Fermi energy $E_f$.
\begin{equation}
    f(E) = \frac{1}{\text{exp}\left(\frac{E-E_f} T\right)+1}
\end{equation}
represents the fraction of occupied states at energy $E$.
The scattering cross section $\frac{d\sigma}{d\Omega}$ is analogous to a Rutherford scattering cross section \cite{PhysRevD.26.2347}
\begin{equation}
    \frac{d\sigma}{d\Omega} = \frac{Q^2e^2}{4p^2\text{sin}^4(\Psi/2)},
\end{equation}
where $\text{cos}\Psi = \text{sin}\theta\text{sin}\theta'\text{cos}\Delta\phi+\text{cos}\theta\text{cos}\theta'$.  $\Psi$ is the angle between the velocities of the incoming and scattered electron.  The integral over $\Psi$ diverges as a log as $\Psi$ approaches zero.  We cut it off at $\Psi_{\text{min}} = \frac{\nu_{\sigma}^{ie}}{2E_f}$, which limits the scattering angle by the scattering length in the medium \cite{PhysRevD.26.2347}.  Integrating Eq.\eqref{startingfric} over $\Psi$ and $\phi$ gives $16\pi \log\left(\frac{1}{\Psi}\right)$.  Substituting $x = \cos\theta$ and $y =\cos\theta'$ gives:
\begin{equation}
    \frac{dE}{dx}=\frac{Ge^2\sqrt{2m_e}M^2}{\pi^2}\text{log}\left[\frac{2E_f}{\nu_{\sigma}^{ei}}\right]\times\int_{-1}^{1}\int_{-1}^{1}\int_0^{\infty}\sqrt{E}\times f(E)\left(1-f(E+\Delta E)\right)
    dx dy dE,
\end{equation}
where $G$ is the gravitational constant, $e$ is the electron charge, $m_e$ is the elctron mass, and $\nu_{\sigma}^{ei}\sim 10^{17}$ s$^{-1}$ is the electron ion scattering frequency \cite{Potekhin:1999yv}.
\section{Calculation Details for Binary Mergers}
\subsection{Derivation of the Merger Time for an EMBH Binary}
\label{merger time d}
The time it takes an EMBH binary to merge can be found by considering the energy lost to EM radiation emitted during orbit.  EMBHs moving through an orbit radiate as two perpendicular oscillating magnetic dipoles.  Energy losses via gravitational radiation come from the binary's quadropole moment, and are subdominant.  The binary emission rate for black holes of arbitrary electric charge and mass in circular orbits is presented in \cite{Liu:2020vsy,Liu_2020}.  The emission rate for black holes with arbitrary electric and magnetic charge can be found in \cite{liu2020gravitational}.  The merger rate for charged black holes in a circular orbit is also considered in  \cite{ghosh2020astrophysical}. 

The electric and magnetic fields of two perpendicular magnetic dipoles which vary arbitrarily with time can be described in spherical coordinates by orienting them in the $\theta = \pi/2$ plane.
\begin{equation}
\label{EMfield}
\begin{split}
    B = & \frac{2\cos\theta}{r^2}\left[\frac{1}{r}+\frac{d}{d\tau}\right]\left(m_2(\tau)+m_1(\tau)\right)\hat{r}\\
    &+\frac{1}{r}\left[\frac{1}{r^2}+\frac{1}{r}\frac{d}{d\tau}+\frac{d^2}{d\tau^2}\right]\left(-\sin\theta m_1(\tau)+\cos\theta\cos\phi m_2(\tau)\right) \hat{\theta}\\
    & -\frac{\sin\phi}{r}\left[\frac{1}{r^2}+\frac{1}{r}\frac{d}{d\tau}+\frac{d^2}{d\tau^2}\right]m_2(\tau)\hat{\phi}\\
    \\
    E = &-\frac{\sin\phi}{r}\left[\frac{1}{r}\frac{d}{d\tau}+\frac{d^2}{d\tau^2}\right]m_2(\tau)\\
    &+\frac{1}{r}\left[\frac{1}{r}\frac{d}{d\tau}+\frac{d^2}{d\tau^2}\right]\left(\sin\theta m_1(\tau)+\cos\theta\cos\phi m_2\tau\right)\hat{\phi},
\end{split}
\end{equation}
where $m_1(\tau)$ and $m_2(\tau)$ are the dipole moments as a function of the retarded time $\tau = t-r$, and $r$ is the radius from the center of the dipole. The energy loss is found by taking $\frac{dE}{dt} =\int \frac{1}{4\pi}(E\times B)\cdot dA$  at $r=\infty$.  Changes to the shape of the orbit are described by changes to the angular momentum along the $\theta=0$ axis, which can be described by $\frac{dL}{dt}=\frac{1}{4\pi}\int \cos\theta\left(r\times\left(E\times B\right)\right)dA$ integrated over a sphere at $r=\infty$. Using the terms from Eq.(\ref{EMfield}), this gives:
\begin{equation}
\label{angloss}
\begin{split}
    \frac{dE}{dt}& =\frac{2}{3}\left[\left(\frac{d^2}{d\tau^2}m_2(\tau)\right)^2+\left(\frac{d^2}{d\tau^2}m_1(\tau)\right)^2\right]\\
    \\
    \frac{dL}{dt}& =\frac{2}{3}\left[\frac{d^2m_2}{d\tau^2}\frac{dm_1}{d\tau}+ \frac{d^2m_1}{d\tau^2}\frac{dm_2}{d\tau}\right].
\end{split}
\end{equation}

To use these expressions to get a merger time, one must describe the dipole moments in terms of orbital parameters of the binary system.  Each EMBH moves about the center of mass of the system with an angular velocity:
\begin{equation}
    \Dot{\phi}=\sqrt{\frac{2G(M_1+M_2)}{a^3(1-e^2)^3}}(1+e\cos\phi)^2,
\end{equation}
where $a$ is the semi-major axis of the orbit, $e$ is the eccentricity, and $M_1$ and $M_2$ are the masses of each EMBH.  EMBHs feel an equal mutual magnetic and gravitational attractive force, which is accounted for by the extra factor of 2 in the above expression.  The distance of an EMBH from the center of mass of the system, $d$, is set by:
\begin{equation}
    d = \frac{a(1-e^2)}{1+e\cos\phi}.
\end{equation}
In terms of these parameters the dipole moments are:
\begin{equation}\begin{split}
    m_1(\tau)& = 2\frac{M_1}{m_p} d(\tau)\cos\phi\\
    \\
    m_2(\tau)& = 2\frac{M_1}{m_p}d(\tau)\sin\phi,
\end{split}
\end{equation}
where $m_p$ is the Plank mass and $m_{P}^{-2}=G$.  Using these expressions, all of the time derivatives of the dipole moments can be found in terms of $a$, $e$, and $\phi$.  Putting these into Eq.(\ref{angloss}) gives:
\begin{equation}
\begin{split}
     \frac{dE}{dt} &= \frac{32G^3 M_1M_2(M_1+M_2)^2}{3a^4(1-e^2)^4}(1+e\cos\phi)^4\\
     \\
     \frac{dL}{dt} &=\frac{8G^{5/2}M_1M_2\left(2\left(M_1+M_2\right)\right)^{3/2}}{3a^{5/2}(1-e^2)^{5/2}}(1+e\cos\phi)^3.
\end{split}
\end{equation}
Averaging the energy and angular momentum losses over a single orbit gives:
\begin{equation}
    \begin{split}
        \left\langle \frac{dE}{dt}\right\rangle& = -\frac{4G^3M_1M_2(M_1+M_2)^2}{a^4(1-e^2)^4}\left(\frac{8}{3}+8e^2+e^4\right)\\
        \\
        \left\langle\frac{dL}{dt}\right\rangle& = -\frac{G^{5/2}M_1M_2(2(M_1+M_2))^{3/2}}{a^{5/2}(1-e^2)^{5/2}}\left(\frac{8}{3}+4e^2\right).
    \end{split}
\end{equation}

For comparison to the energy losses from gravitational radiation, we rewrite these assuming $M_1=M_2=M$.
\begin{equation}
    \begin{split}
       \left\langle\frac{dE}{dt}\right\rangle& = -\frac{16M^4G^3}{a^4(1-e^2)^4}\left(\frac{8}{3}+8e^2+e^4\right) \\
        \\
        \left\langle\frac{dL_y}{dt}\right\rangle& = -\frac{M^{7/2}2^{3/2}G^{5/2}}{a^{5/2}(1-e^2)^{5/2}}\left(\frac{8}{3}+4e^2\right)
    \end{split}
\end{equation}
The power output from gravitational waves for an EMBH binary  is \\$ \left\langle \frac{dE}{dt}\right\rangle =\frac{1024}{5}\frac{G^4M^5}{a^5(1-e^2)^{7/2}}\left(1+\frac{73}{24}e^2+\frac{37}{96}e^4\right)$ \cite{PhysRev.131.435}, very small compared to the power from EM radiation.  The extra factor of 16 comes from enhancements to the energy and angular momentum of the binary due to the magnetic attraction between EMBHs.   Power loss from gravitational radiation will only match EM radiation once $a$ drops down to $\frac{12}{5}r_s$, where $r_s=2GM$ is the Schwarzschild radius, and is thus unimportant for estimating the time needed for EMBH binaries to merge. 

The energy of the system is related to the semi-major axis by $E = -\frac{GM^2}{a}$.  This is twice the energy of a neutral black hole system of the same mass due to the extra attractive force between the EMBHs.  The average rate of energy loss can be related to the average change in the semi-major axis by $\left\langle \frac{dE}{dt}\right\rangle = \frac{GM^2}{a^2}\left\langle\frac{da}{dt}\right\rangle$, which gives;
\begin{equation}
\label{EQ:DADT}
    \left\langle\frac{da}{dt}\right\rangle = \frac{16G^2M^2}{a^2(1-e^2)^4}\left(\frac{8}{3}+8e^2+e^4\right).
\end{equation}
For $e=0$ one can solve for the merger time analytically
\begin{equation}
    t_{m0} = \frac{a_0^3}{128 G^2M^2} = 4.8\times10^{49}s \frac{(a_0/\text{m})^3}{(M/\text{g})^2},
\end{equation}
where $a_0$ is the initial semi-major axis.  For comparison, the merger time if only the gravitational radiation were considered would be 
\begin{equation}
    t_{Gm0} = \frac{5}{512\sqrt{2}}\frac{a^4}{G^3M^3} = 5.8\times10^{79}s \frac{(a_0/\text{m})^4}{(M/g)^3}.
\end{equation}

When $e\neq 0$ the merger time also depends on how the eccentricity evolves.  Using $L^2 = G M^3a(1-e^2)$ gives the relation
\begin{equation}
    L\left\langle\frac{dL}{dt}\right\rangle=GM^3\left((1-e^2)\left\langle\frac{da}{dt}\right\rangle-2ae\left\langle\frac{de}{dt}\right\rangle\right),
\end{equation}
which can be rearranged to find
\begin{equation}
\label{eq:dedt1}
    \left\langle\frac{de}{dt}\right\rangle = \frac{8G^2M^2}{a^3(1-e^2)^3}\left(\frac{20}{3}e+5e^3\right).
\end{equation}
Eqs.\eqref{EQ:DADT} and \eqref{eq:dedt1} can be used to find
\begin{equation}
\label{dade}
   \left\langle\frac{da}{de}\right\rangle = \frac{2a}{(1-e^2)}\frac{\left(\frac{8}{3}+8e^2+e^4\right)}{\left(\frac{20}{3}e+5e^3\right)}. 
\end{equation}
Integrating this gives $a$ as a function of $e$:
\begin{equation}
\label{aofe}
    a = a_0\frac{(1-e_0^2)}{e_0^{4/5}(3e_0^2+4)^{2/5}}\frac{e^{4/5}(3e^2+4)^{2/5}}{(1-e^2)},
\end{equation}
where $a_0$ and $e_0$ are the starting semi-major axes and eccentricities of the system.  Finally, Eq.\eqref{aofe} can be plugged into Eq.\eqref{eq:dedt1} and integrated to get a general expression of the merger time as a function of the initial semi-major axis and eccentricity of a binary system.
\begin{equation}
    \label{mergetime}
    t_{m} = \frac{a_0^3}{8G^2M^2}\frac{ (1-e_0^2)^3}{e_0^{12/5}(3e_0^2+4)^{6/5}}\int_0^{e_0}e'^{7/5}\frac{(3e'^2+4)^{6/5}}{(\frac{20}{3}+5e'^2)}de'
\end{equation}
In the limit $e_0\sim 1$, the integral over $e$ becomes $\sim 0.35$, and $t_m$ becomes
\begin{equation}
\label{mergetimehe}
    t_m = \frac{0.034a_0^3}{8G^2M^2}(1-e_0^2)^3.
\end{equation}
If we do not assume $M_1=M_2=M$ then this is
\begin{equation}\label{t2}
    t_m = \frac{0.034a_0^3}{2G^2(M_1+M_2)^2}(1-e_0^2)^3.
\end{equation}

\subsection{Derivation of the EMBH Binary Merger Rate }
\label{sec: app merge}
The EMBH merger rate can be found by modifying derivations for the merger rate of uncharged black hole binaries  \cite{Sasaki:2018dmp,Nakamura:1997sm} to account for the extremal magnetic fields.  The initial orbital parameters of binaries, their semi-major axes, $a_0$, and eccentricities, $e_0$, are characterized by their co-moving separation, $x$, the co-moving distance to the next nearest EMBH, $y$, and the redshift, $z_d$, when the binary decouples from the Hubble flow.  Co-moving distances will correspond to distances in the present, when the scale factor equals 1, if the separated objects never decouple from the Hubble flow.

When friction with the thermal bath is not important, EMBH binaries decouple from the Hubble flow and form a stable binary when the time needed for them to free-fall into each other becomes less than the Hubble time \cite{Sasaki:2018dmp,Nakamura:1997sm}.  During radiation domination this corresponds to a decoupling redshift
\begin{equation}
\label{zd}
     (1+z_{d}) = \frac{3M(1+z_{eq})}{4\pi \rho_{dm} x^3} = \frac{k_dM}{x^3},
\end{equation}
where $k_d\equiv \frac{3(1+z_{eq})}{4\pi \rho_{dm}}$, $\rho_{dm}$ is the dark matter density, and $z_{eq}$ is the redshift of matter radiation equality. 

The infall dynamics change when friction with the thermal bath becomes large compared to Hubble acceleration.  In this scenario, the binary decouples when the EMBH terminal infall velocity exceeds the Hubble velocity $v_H = \frac{H(z)x}{(1+z)}$, where $H(z)$ is the Hubble expansion rate at a redshift $z$.  The terminal infall velocity is found by balancing the attractive force between the EMBHs,  $\frac{2GM^2(1+z)^2}{x^2}$, with the frictional force exerted by the plasma on them.  The frictional force between a magnetic monopole and a plasma is described in Ref.\cite{1985ApJ...290...21M}. We adapt it here for an EMBH with mass $M$:
\begin{equation}
    \left(\frac{dE}{dx}\right)_f = -\frac{16 \pi^{1/2}e^2n_e(z)}{3\sqrt{2T(z) m_e}}G M^2 V \left[\log\left( 4\pi n_e(z) \lambda_D^2l\right)+\frac{2}{3}\right],
\end{equation}
where $n_e$, is the electron number density in the thermal bath which we take to be $\sim(1+z)^3\frac{\rho_{c0} \Omega_b}{m_{pr}}$,  $\rho_{c0}$ is the critical density today, $\Omega_b$ is the baryon fraction, $m_{pr}$ is the proton mass, $e$ and $m_e$ are the electron charge and mass, respectively, $T$ is the gas temperature which we take to be $\sim 2.7(1+z)$K, $V$ is the velocity of the EMBH, $\lambda_D=\sqrt{\frac{T}{4\pi n_e e^2}}$ is the plasma Debye length, $l = \left(\frac{2T}{\pi m_e}\right)^{1/4}\frac{1}{V^{1/2}\omega_p}$ is the attenuation length of the plasma and $\omega_p$ is the plasma frequency.  To simplify comparing forces we will represent $(\frac{dE}{dx})_f$ as $M^2V(1+z)^{5/2}\mathcal{F}$.  $\mathcal{F}$ represents the \textit{mostly}  $z$, $M$ and $V$ independent component of $(\frac{dE}{dx})_f$

If $\mathcal{F}z^{5/2}M>H(z) $ at the regular decoupling redshift, $z_d$, then the EMBHs will continue moving apart with the Hubble flow until $\frac{2G}{x^2(1+z)^{1/2}\mathcal{F}} = \frac{H(z)x}{(1+z)}$.  During radiation domination this gives a decoupling redshift 

\begin{equation}
\label{eqn:zb}
    (1+z_b) = \left(\frac{2G}{H_0\sqrt{\Omega_r}\mathcal{F}}\right)^{2/3}x^{-2} = k_b x^{-2},
\end{equation}
where $k_b\equiv\left(\frac{H_0\sqrt{\Omega_r}\mathcal{F}}{2G}\right)^{2/3}$, $\Omega_r$ is the fraction of the critical energy density composed of radiation at $z=0$, and $H_0$ is the Hubble constant.  Binaries must decouple by matter radiation equality because the ratio between the free-fall time and the Hubble time becomes constant during matter domination.

The semi-major axis of the binary is $a_0=\frac{x}{2(1+z_d)}$.
The eccentricity, $e_0$, of the binary is set by interactions with the next nearest EMBH.   It pulls the opposite charge EMBH in the binary out of it's free-fall path, without affecting the same charge one. The displacement from its free-fall path sets the semi-minor axis for the binary, $b$.  This can be estimated by 
\begin{equation}
    2b = \frac{1}{2}\left(\frac{2GM(1+z_d)^2}{y^2}\right)\left(\frac{x^3}{2GM(1+z_d)^3}\right)=\frac{x^2}{y^2}a,
\end{equation}
which gives a starting eccentricity 
\begin{equation}
\label{eccentricity}
    e_0 = \sqrt{1-\frac{x^4}{4y^4}}.
\end{equation}

The probability that two oppositely charged EMBHs are separated by a co-moving distance between $x$ and $x+dx$ and that the next nearest one is between $y$ and $y+dy$ can be described by a Poisson distribution.
\begin{equation}
\label{eq:startprob}
    dp = \frac{1}{2}(4\pi n)^2x^2y^2dxdy,
\end{equation}
where $n$ is the co-moving number density of EMBHs.  The factor of $\frac{1}{2}$ arises because each EMBH can only pair with an oppositely charged partner.  $x$ is by definition less than $y$. $y$ can be at most the average separation between EMBHs.  This defines
\begin{equation}
    y_{\text{max}}\equiv \left(\frac{3M}{4\pi \rho_{dm}f}\right)^{1/3}.
\end{equation}
Eq.\eqref{eq:startprob} can be re-written using Eqs.\eqref{zd} and \eqref{eccentricity} to give a probability distribution for the orbital parameters of all of the binaries.  That can then be related to the probability that a binary merges at a specific time, $t_m$, by Eq.\eqref{mergetimehe} which is derived in Section \ref{merger time d}.  
Rearranging Eq.\eqref{mergetimehe} gives $a_0(t_m,e_0)$, the starting semi-major axis that corresponds to a binary with eccentricity $e_0$ that would merge in $t_m$.
\begin{equation}
    a_0 = \frac{ M^{2/3}t_m^{1/3}}{\left(\frac{0.034}{8G^2}\right)^{1/3}(1-e_0^2)}
\end{equation}
All binaries that merge at $t_m$ have parameters that lie along the curve in $e-a$ space $a_0(t_m,e_0)$.
With a bit of algebra, one finds when $e\sim 1$ the probability that an EMBH binary merges between $t_m$ and $t_m + dt$ is 
\begin{equation}
\label{dpdtde}
    dp = \frac{(4\pi n)^2}{24}\frac{M^{5/2}k_d^{3/2}}{\left(\frac{0.034}{8G^2} t_m\right)^{1/2}}\frac{e_0}{(1-e_0^2)^{13/4}}dtde
\end{equation}
The merger rate per EMBH, $\Gamma_m$, at a specific $t_{m}$ can be found by integrating $\frac{dp}{dt}$ over all possible starting eccentricities for a binary merging at $t_{m}$.  
\begin{equation}
\label{eqn:ratemerge}
    \Gamma_m=\frac{dp}{dt} = \frac{(4\pi \rho_{dm} f)^2}{27*4}\frac{M^{1/2}k_d^{3/2}}{\left(\frac{0.034}{8G^2}t_m\right)^{1/2}}\left(\frac{1}{(1-e_{\text{max}}^2)^{9/4}}-\frac{1}{(1-e_{\text{min}}^2)^{9/4}}\right)
\end{equation}
$f$ is the fraction of dark matter composed of EMBHs.  Multiplying $\Gamma_m$ by $n_{bh}$, the proper number density of EMBHs at $t_m$, would give the rate at which EMBH mergers happen at $t_m$. 

The bounds used in these integrals $e_{\text{max}}$ and $e_{\text{min}}$ are the largest and smallest eccentricities a binary merging at $t_m$ could have.  
We define $e_{upper}$ as the eccentricity a binary with semi-major axis $a_0$  and $y=y_{\text{max}}$, the highest eccentricity possible for that binary given the average separation between EMBHs.  
\begin{equation}
    e_{upper}=\sqrt{1-\frac{a_0}{4k_dy_{\text{max}}^4}}
\end{equation}
$e_{\text{max}}$ is set by where the $a_0(e_0,t_m)$ curve intersects $e_{upper}$ or $a_{eq}$, the semi-major axis of a binary that decouples at the latest allowed decoupling redshift $z_{eq}$.
\begin{equation}
e_{\text{max}}=
 \begin{cases}
\sqrt{1-\left(\frac{ M^{2/3}t_m^{1/3}}{4k_d\left(\frac{0.034}{8G^2}\right)^{1/3}y_m^4}\right)^{1/2}} & t_m\leq T_c\\
\sqrt{1-\left(\frac{ k_d M^2t_m(1+z_{eq})^4}{\left(\frac{0.034}{8G^2}\right)}\right)^{1/3}} & t_m > T_c,
\end{cases}   
\end{equation}
where
\begin{equation}
    T_c=\frac{\frac{0.034}{8G^2}}{2^6M^2k_d^5y_m^{12}(1+z_{eq})^8}
\end{equation}  Finally, $e_{\text{min}}$ is the lowest eccentricity a binary that merges in $t_m$ and decouples after $z_f$ can have.  
$z_f$ denotes the redshift at which frictional and Hubble acceleration are equal:
\begin{equation}
    (1+z_f) = \frac{\mathcal{F}M^2}{H_0\sqrt{\Omega_r}}.
\end{equation}
$e_{\text{min}}$ is 0 when all binaries that merge in $t_m$ decouple after $z_f$. 
\begin{equation}
    e_{\text{min}} =
    \begin{cases}
    \sqrt{1-\left(\frac{ k_d M^2t_m(1+z_{f})^4}{\left(\frac{0.034}{8G^2}\right)}\right)^{1/3}} & t_m > T_f\\
    0 & t_m < T_f,
    \end{cases}
\end{equation}
where  
\begin{equation}
    T_f = \frac{\left(\frac{0.034}{8G^2}\right)}{ M^2k_d(1+z_{f})^4}.
\end{equation}
Here $e_{\text{min}}$ is determined in the limit that $e_{\text{min}}\sim1$ because the merger rate is dominated by binaries with $e\sim1$.

If one wanted to generalize the merger rate to EMBHs with a non-monochromatic mass specturm then one could follow the same general procedure outlined above with some modified values of $dp$, $z_d$, and $e$.  Instead of simply integrating over the range of eccentricities that will lead a binary to merge in the present, one must integrate over all possible eccentricities and EMBH masses for a given mass distribution.  One can approach this by rewriting Eq.\eqref{eq:startprob} as
\begin{equation}
\label{p2}
    dp = \frac{1}{2}(4\pi n)^2x^2y^2\delta(t_m-t)\chi (M_1)\chi(M_2)\chi(M_3)dxdydM_1dM_2dM_3
\end{equation}
where $\chi(M)$ is the mass pdf such that $\int\chi(M)dM = 1$, $M_1$ and $M_2$ are the masses of the EMBHs in the binary, and $M_3$ is the mass of the third EMBH that perturbs the infalling pair and sets the eccentricity, and $t$ is the time the mergers take place.  $t_m$ is given in Eq.\eqref{t2} and a function of $a_0$, $e_0$, $M_1$, and $M_2$.

The decoupling redshift for EMBHs of different masses is
\begin{equation}
\label{z2}
    (1+z_d) = \frac{6M_1M_2(1+z_{eq})}{4\pi \rho_{dm}(M_2+M_2) x^3}.
\end{equation}
The generalized eccentricity is
\begin{equation}
\label{e2}
      e_0 = \sqrt{1-\frac{x^4}{4y^4}\left(\frac{M_3(M_1+M_2)}{M_1M_2}\right)^2}.
\end{equation}
Substituting Eqs. \eqref{z2}, \eqref{e2}, and \eqref{t2} into Eq.\eqref{p2} in the same manner as was done above will give the probability that a binary has the correct eccentricity and masses to merge in $t_m$.  One can integrate this over the range of possible eccentricities, noting that $y_{max} = \left(\frac{3\langle M\rangle}{4\pi \rho_{dm}f}\right)^{1/3}$, and the range of possible masses.  This will likely need to be done numerically for arbitray mass distributions. 

\bibliographystyle{JHEP}
\bibliography{monopolebib}

\end{document}